\documentclass[structabstract]{aa}  
\usepackage{graphicx}
\usepackage{natbib}                                  
\usepackage{lscape}
\usepackage{txfonts}
\def\ltr{$L_{\rm X}-T$ }

\begin{document}

   \title{The 2XMMi/SDSS Galaxy Cluster Survey}

   \subtitle{II. The optically confirmed cluster sample and the \ltr relation\thanks{Full Tables 1 and 2 are only available in electronic form
at the CDS via anonymous ftp to cdsarc.u-strasbg.fr (130.79.128.5)
or via http://cdsweb.u-strasbg.fr/cgi-bin/qcat?J/A+A/.../.... } }

   \author{A. Takey \inst{1,2}, 
           A. Schwope\inst{1}, \and
           G. Lamer\inst{1}
           }

   \institute{Leibniz-Institut f{\"u}r Astrophysik Potsdam (AIP),
              An der Sternwarte 16, 14482 Potsdam, Germany\\
              \email{atakey@aip.de}
         \and
             National Research Institute of Astronomy and Geophysics (NRIAG), 
             Helwan, Cairo, Egypt             
             }

    \date{Received ....; accepted ....}



   \abstract
   {}
   {We compile a sample of X-ray-selected galaxy groups and clusters 
 from the XMM-Newton serendipitous source catalogue (2XMMi-DR3) with optical 
confirmation and redshift measurement from the Sloan Digital Sky Survey (SDSS).
We present an analysis of the X-ray properties of this new sample with
particular emphasis on the X-ray luminosity-temperature (\ltr) relation.}
   {The X-ray cluster candidates were selected from the 2XMMi-DR3 catalogue in 
the footprint of the SDSS-DR7. We developed a finding algorithm to search for 
overdensities of galaxies at the positions of the X-ray cluster candidates in 
the photometric redshift space and to measure the redshifts of the clusters 
from the SDSS data. 
For optically confirmed clusters with good quality X-ray data we derived
the X-ray flux, luminosity,  and temperature from proper spectral fits, while
the X-ray flux for clusters with low-quality X-ray data was obtained from the
2XMMi-DR3 catalogue.
}
   {The detection algorithm provides the photometric redshift of 530 
galaxy clusters. Of these, 310 clusters have a spectroscopic redshift for at 
least one member galaxy. About 75 percent of the optically confirmed cluster 
sample are newly discovered X-ray clusters. Moreover, 301 systems are known as optically 
selected clusters in the literature while the remainder are new discoveries in 
X-ray and optical bands. The optically confirmed cluster sample spans a wide 
redshift range 0.03-0.70 (median $z$=0.32). 
In this paper, we present the catalogue of X-ray-selected galaxy groups and 
clusters from the 2XMMi/SDSS galaxy cluster survey. The catalogue has two 
subsamples: (i) a cluster sample comprising 345 objects with their X-ray 
spectroscopic temperature and flux from the spectral fitting, and 
(ii) a cluster sample consisting of 185 systems with their X-ray flux from the 
2XMMi-DR3 catalogue, because their X-ray data are insufficient for spectral 
fitting. For each cluster, the catalogue also provides the X-ray bolometric 
luminosity and the cluster mass at $R_{500}$ based on scaling relations and 
the position of the likely brightest cluster galaxy (BCG). 
The updated \ltr relation of the current sample with X-ray spectroscopic 
parameters is presented. We found the slope of the \ltr relation to be consistent
with published ones. We see no evidence for evolution in the slope and 
intrinsic scatter of the \ltr relation with redshift when excluding the 
low-luminosity groups.
}
   {}


   \maketitle
%


\section{Introduction}

   Galaxy clusters are the largest known gravitationally bound objects;  
studying them is important for both an intrinsic understanding of their systems and  
an investigation of the large-scale structure of the Universe. The multi-component 
nature of galaxy clusters offers multiple observable signals across the 
electromagnetic spectrum \citep[e.g.][]{Sarazin88, Rosati02}. 
The hot, ionised intra-cluster medium (ICM) is investigated  at X-ray 
wavelengths and using the Sunyaev-Zeldovich (SZ) effect 
\citep[][]{Sunyaev72,Sunyaev80}. The cluster galaxies are most effectively 
studied through  optical and near-infrared (NIR) photometric and spectroscopic 
surveys. The statistical studies of galaxy clusters provide complementary 
and powerful constraints on the cosmological parameters \citep[e.g.][]{Voit05, 
Allen11}.
 
X-ray observations offer the most powerful technique for constructing cluster
catalogues. The main advantages of the X-ray cluster surveys are their excellent 
purity and completeness, and moreover, the X-ray observables are tightly 
correlated with mass \citep[e.g.][]{Reiprich02, Allen11}. 
Reliable measurements of cluster masses allow us to measure 
both the mass function \citep{Boehringer02} and the power spectrum 
\citep{Schuecker03}, which directly probe the cosmological models. 

At X-ray wavelengths, galaxy clusters are simply identified as X-ray luminous, 
continuous, spatially extended, extragalactic sources \citep{Allen11}. 
Several X-ray cluster samples have been constructed from previous X-ray 
missions and have been used for a variety of 
astrophysical studies \citep[e.g.][]{Romer94, Forman78, Scharf97, Vikhlinin98, 
Boehringer00, Borgani01, Boehringer04, Burenin07}. The current generation of 
X-ray satellites, XMM-Newton, Chandra, and Suzaku, provided follow-up observations 
of individual clusters that allowed a precise determination of their spatially 
resolved spectra \citep[e.g.][]{Vikhlinin09,Pratt10,Arnaud10}. Several other 
projects are being conducted to detect galaxy clusters from the observations of 
the XMM-Newton, Chandra, and the X-ray Telescope on board of the Swift 
satellite \citep[e.g.][]{Barkhouse06, Kolokotronis06, Finoguenov07, 
Finoguenov10, Adami11, Fassbender11, Takey11, Mehrtens12, Clerc12, 
Tundo12, de-Hoon13}.

We have started a serendipitous search for galaxy clusters based on extended 
sources in the 2XMMi-DR3 catalogue, the second XMM-Newton source 
catalogue \citep{Watson09}, in the footprint of the SDSS-DR7. The main aim of 
the survey is to construct a large catalogue of newly discovered X-ray-selected 
groups and clusters from XMM-Newton archival observations. The catalogue will 
allow us to investigate the evolution of X-ray scaling relations as well as 
the correlation between the X-ray and optical properties of the clusters. 

The survey comprises 1180 X-ray-selected cluster candidates. A cross-correlation 
of these with recently published optically selected SDSS galaxy cluster 
catalogues yielded photometric redshifts for 275 objects. 
Of these, 175 clusters were published by \citet[][paper I hereafter]{Takey11} 
together with their X-ray luminosity, temperature, and mass.
The first cluster sample from the survey covers a wide range of redshifts from 
0.09 to 0.61.
We extended the relation between the X-ray bolometric 
luminosity at $R_{500}$  (the radius at which the cluster mean density 
is 500 times the critical density of the Universe at the cluster redshift)
and the X-ray temperature towards significantly lower 
luminosities than reported in the literature and found that the slope of the
linear \ltr relation was consistent with that for more luminous clusters.

In the present paper, we expand the optically confirmed sample from the survey 
by searching for the optical counterparts of cluster candidates that had been 
missed by previous cluster-finding algorithms and their members detected in the 
SDSS imaging (see paper I for a sample of X-ray and optically selected groups 
and clusters).
We present the algorithm we used to identify the optical counterparts of the 
X-ray cluster candidates and to estimate the cluster redshifts using SDSS data. 
As a result, we present a catalogue of X-ray-selected galaxy groups and clusters
(including the objects in paper I) from the ongoing 
2XMMi/SDSS galaxy cluster survey.
The catalogue provides the X-ray properties (such as temperature, flux, luminosity, 
and mass) and the cluster photometric redshift and, where available, the 
cluster spectroscopic redshift and the position of the likely brightest cluster 
galaxy (BCG) of the optically confirmed cluster sample. 

The X-ray luminosity-temperature (\ltr) relation was investigated by several 
authors \citep[e.g.][]{Markevitch98, Pratt09, Mittal11, Eckmiller11,  Reichert11, 
Takey11, Maughan12, Hilton12}. 
These studies showed that the observed \ltr relation is much steeper than that  
predicted by self-similar evolution. This indicates that the ICM 
is heated by an additional source of energy, which is mainly active 
galactic nuclei (AGN) \citep{Blanton11}. 
Including of AGN-feedback in cosmological evolution models indeed gives
a better agreement between simulated and observed \ltr under certain
circumstances \citep{Hilton12}.
Here, we present an updated \ltr relation based on the largest sample of 
X-ray-selected groups and clusters to date drawn from a single survey based on 
XMM-Newton observations. The sample spans a wide redshift range from 
0.03 to 0.67. 

The organisation of this paper is as follows: In Section 2, we describe  
the construction of the X-ray cluster candidate list and the optically 
confirmed cluster sample with their redshift estimates. In Section 3, 
we present the X-ray data reduction and analysis of the sample. In 
Section 4, the results and discussion of the cluster sample is 
presented. We summarise our results in Section 5. The cosmological 
parameters  $\Omega_{\rm M}=0.3$, $\Omega_{\Lambda}=0.7$, and 
$H_0=70$\ km\ s$^{-1}$\ Mpc$^{-1}$ were used throughout this paper.



\section{Sample construction}

We started our search based on the XMM-Newton 
serendipitous sources followed by searching for overdensities of galaxies in 
3D space. In the following subsections, we present the strategy we followed to 
create the X-ray cluster candidate list. To derive the X-ray properties of these 
candidates, we needed to determine their redshift either from the X-ray data, 
which is only possible for the X-ray-brightest clusters, or from the optical 
data, which is the way we used in the current work. 
We also present the algorithm we used to detect the clusters in the optical 
band and to estimate their redshifts from the SDSS data. We compare the 
measured redshifts with the published values.


\subsection{X-ray cluster candidates}

The survey comprises X-ray cluster candidates selected as serendipitous 
sources from the 2XMMi-DR3 catalogue in the footprint of SDSS-DR7.
The number of XMM-Newton fields that were used in constructing the 
2XMMi-DR3 catalogue in the footprint of SDSS-DR7 at high galactic latitude 
$|b| > 20\degr$ is 1200 fields after excluding the multiple observations 
of the same field. We also excluded fields that were flagged as bad 
(the whole field) and unsuitable for source detection according to the  
manual flag given in the 2XMMi-DR3 catalogue. The total area of the fields 
included in our survey is 210 deg$^2$, taking into account the overlap areas 
among the fields.  

The cluster candidate selection was based on X-ray-extended sources that 
passed the quality assessment during the construction of the
catalogue  by the XMM-Newton Survey Science Center (SSC).
The extent parameter of each extended source in the 2XMMi-DR3 catalogue is 
determined by the SAS task {\tt emldetect} by fitting a convolution of 
a $\beta$ model ($\beta$=2/3) and the instrument point spread function (PSF) 
to each input source. The source is classified as extended if the extent 
parameter varies between 6 to 80 arcsec and if the extent likelihood is 
higher than 4 \citep{Watson09}.

The completeness of the 2XMMi-DR3 extended source catalogue is not easy to 
assess because the 2XMMi-DR3 catalogue was constructed from 4953 
observations with different exposure times. The wide range of exposure 
times yields various flux limits.  
\cite{Muehlegger10} simulated two fields (LBQS and SCSA with exposure time 
52 ks and 8.8 ks, respectively) in the XMM-Newton Distant Cluster project 
(XDCP) to test the detection probability. They used a source 
detection technique that is similar to the one used in detecting the 2XMMi-DR3
sources. 
According to their simulations, the higher detection probability was achieved  
for clusters with intermediate core radii in the range of 15 to 25 arcsec. The
probability decreases with decreasing photon counts and decreasing  
core radius ($< 7$ arcsec) due to the difficulty to distinguish extended sources 
from point sources. The detection probability of sources with large core radii 
($ > 75 $ arcsec) and a low number of photons was low because these systems 
disappear in the background due to their low surface brightness. 
The detection probability decreases beyond the off-axis angle of 12 arcmin, 
caused by vignetting.
Based on these results, clusters with low photon counts or large core radii 
might be missed in the 2XMMi-catalogue or might be listed with incorrect 
source parameters.

The selected extended sources were visually inspected by us in two steps to  
exclude possibly spurious detections. The first visual inspection was made 
using the X-ray images through the FLIX upper limit 
server\footnote{http://www.ledas.ac.uk/flix/flix.html}. The second one was 
made using the X-ray-optical overlays, where the X-ray flux contours were overlaid 
onto the co-added SDSS images in $r$, $i$, and $z-$bands. 
The first inspection allowed us to remove the obviously spurious cases 
caused by point source confusion, X-ray artefacts, and locations near very bright
sources. Extended sources were also rejected if they were found within another
extended source or at the very edge of the CCDs. The second inspection 
enabled us to also remove the X-ray-extended sources corresponding to low-redshift  
galaxies. The resulting list includes 
1180 X-ray cluster candidates with at least 80 net photon counts. More than 
75 percent are new X-ray detections of galaxy groups and clusters.  

\begin{figure}
  \resizebox{\hsize}{!}{\includegraphics{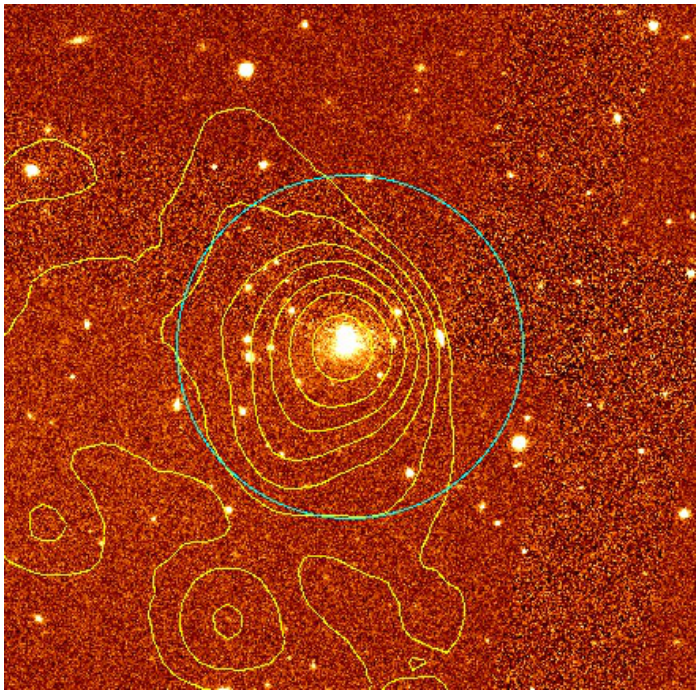}}
  \caption{X-ray-optical overlay of the example cluster
2XMM J102133.2+213752 at spectroscopic redshift = 0.1873. The X-ray flux 
contours (0.2 - 4.5 keV) are overlaid on a combined image from $r$, $i$, and 
$z-$bands SDSS images. The plotted cyan circle has a radius of one arcmin  
around the X-ray emission peak. The field of view is $4'\times 4'$ centred on 
the X-ray cluster position.} 
  \label{f:108143_overlay}
\end{figure}

Figure~\ref{f:108143_overlay} shows the X-ray-optical overlay of a newly 
discovered galaxy cluster in X-ray and optical observations at 
redshift = 0.1873. This cluster has been serendipitously detected (at an off-axis 
angle of about 11 arcmins) in XMM-Newton EPIC observations of the galaxy NGC 3221.
We use this cluster as an example to illustrate the main steps of 
estimating the cluster redshift and the X-ray analyses in the following sections.

\subsection{Constructing the optically confirmed cluster sample}

Various methods have been developed to define the cluster membership of 
galaxies from the data provided by the SDSS. They are based on different 
properties of the clusters and their galaxy members, for instance, using the 
cluster's red sequence, the E/S0 ridge-line 
\citep[e.g.][]{Koester07, Hao10}, or an overdensity of galaxies in the
photometric redshift space \citep{Wen09}. Galaxy clusters are also  
identified by convolving the optical galaxy survey with a set of 
filters in position, magnitude, and redshift space based on modelling 
the cluster and field galaxy distributions \citep{Szabo11}. 

In paper I, we have optically confirmed about a quarter of the X-ray cluster 
candidates through cross-correlation with previously identified clusters in
four optical cluster catalogues \citep[][]{Hao10,Wen09,Koester07,Szabo11}. 
The remainder of the X-ray cluster candidates are either distant cluster 
candidates beyond the SDSS detection limits, that is, z $\ge$ 0.6, which need 
follow-up imaging and spectroscopic confirmation, or there are overdensities 
of galaxies at the X-ray cluster positions that were not recognized by any
previous optical cluster finders (see e.g.~Figure~\ref{f:108143_overlay}).
We therefore developed our own algorithm for clusters with members detected 
in the SDSS imaging to search for optical counterparts and determine their 
redshift from photometric redshifts in the SDSS database.


\subsubsection{Estimation of the cluster redshifts}

Because we have prior information about the cluster position, the position of 
the X-ray emission peak, we can use this information to simplify the 
cluster-finding procedure. We searched for an overdensity of galaxies
around the X-ray position of the cluster candidates within a certain
redshift interval. 
We created a galaxy sample for each X-ray cluster candidate by selecting all
galaxies from the SDSS-DR8 in an area with a radius of 10 arcmin centred on the 
X-ray source position. This radius corresponds to a physical radius of 500\,kpc
at redshift 0.04, which is about our low-redshift limit.

The galaxies were selected from the {\tt galaxy} view table in the SDSS-DR8, which 
contains the photometric parameters measured for resolved primary objects, 
classified as galaxies.  The photometric redshifts and, where available, the 
spectroscopic redshifts of the galaxy sample were also selected from the 
{\tt Photoz} and {\tt Specz} tables, respectively, in the SDSS-DR8. 
The extracted parameters of the galaxy 
sample include the coordinates, the model magnitudes in $r-$band, 
the photometric redshifts, and, where available, the spectroscopic redshifts. 
Where galaxy spectroscopic redshifts were available, we used these 
instead of the photometric redshifts.

\begin{figure}
  \resizebox{\hsize}{!}{\includegraphics{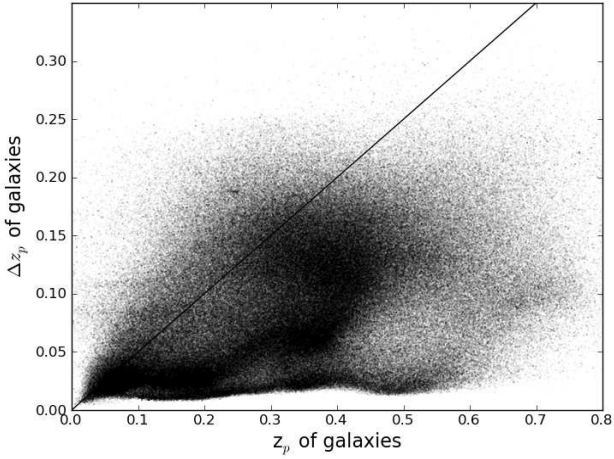}}
  \caption{Photometric redshift error, $ \bigtriangleup z_p $, plotted 
against the photometric redshift, $z_p$, of the galaxy sample with 
$ m_{r} \leq 22$ mag and $\bigtriangleup m_{r} < 0.5 $ mag. The solid line 
indicates the 50 percent relative error of the photometric redshift of the 
galaxy sample.} 
  \label{f:ezp_zp}
\end{figure}

To clean the galaxy sample from faint objects or from galaxies with poor 
photometric measurements, we only used 
galaxies with $ m_{r} \leq 22$ mag and $\bigtriangleup m_{r} < 0.5 $ mag. 
The resulting galaxy sample still includes galaxies with large photometric 
redshift errors, which reach 100 percent in many cases. The photometric 
redshift errors of the galaxy sample with the applied magnitude cut   
are plotted against the photometric redshifts in Figure~\ref{f:ezp_zp}.
To exclude low-redshift galaxies with significantly large relative 
photometric redshift errors as well as to keep high-redshift galaxies 
with moderately large relative errors that were acceptable, we decided 
to apply a relative photometric redshift error cut ($< 50$ percent) instead of 
using a fixed absolute error. The 50 percent relative error line is 
plotted in Figure~\ref{f:ezp_zp}.

\begin{figure}
  \resizebox{\hsize}{!}{\includegraphics{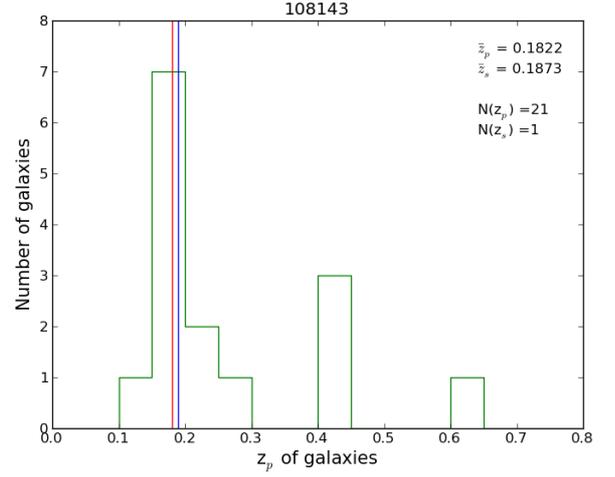}}
  \caption{Photometric redshift distribution of all galaxies within 
one arcmin from the X-ray centroid with $ m_{r} \leq 22$ mag, 
$\bigtriangleup m_{r} < 0.5 $ mag, and $ \bigtriangleup z_{p} / z_{p} < 0.5$. 
The cluster photometric redshift (red line), $\bar{z}_p$, spectroscopic redshift 
(blue line), $\bar{z}_s$, and the cluster member candidates within 560 kpc with 
photometric redshift, N($z_p$), and spectroscopic redshift, N($z_s$), are 
listed in upper right corner.} 
  \label{f:108143_hist1}
\end{figure}

\begin{figure}
  \resizebox{\hsize}{!}{\includegraphics{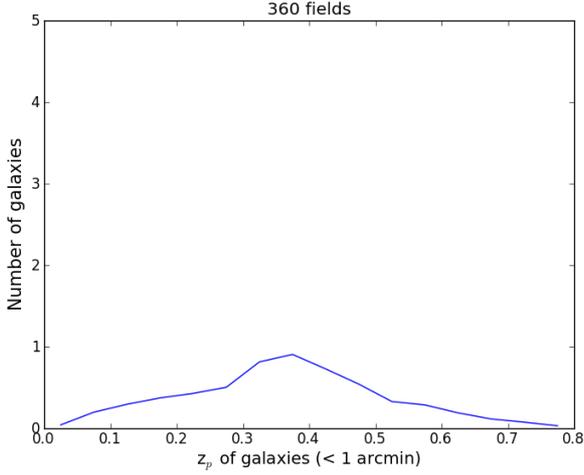}}
  \caption{Distribution of the mean galaxy counts, same distribution as 
Figure~\ref{f:108143_hist1}, within one arcmin from the positions 
of 360 random fields in the SDSS footprint with $ m_{r} \leq 22$ mag, 
$\bigtriangleup m_{r} < 0.5 $ mag, and $ \bigtriangleup z_{p} / z_{p} < 0.5$. 
} 
  \label{f:360_fields}
\end{figure}

\begin{figure}
  \resizebox{\hsize}{!}{\includegraphics{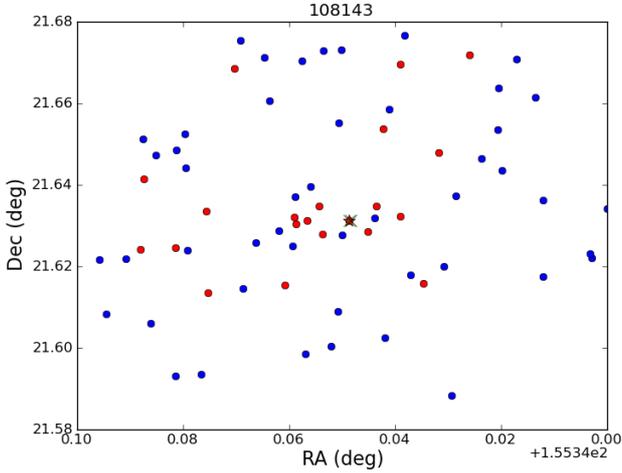}}
  \caption{Sky distribution of cluster galaxies (red dots) and field 
galaxies (blue dots) within 560 kpc from the X-ray position (black X marker ).
The BCG with an available spectroscopic redshift (marked by star) is  
located at the same place as the X-ray cluster position (green x marker).} 
  \label{f:108143_dist}
\end{figure}

\begin{figure}
  \resizebox{\hsize}{!}{\includegraphics{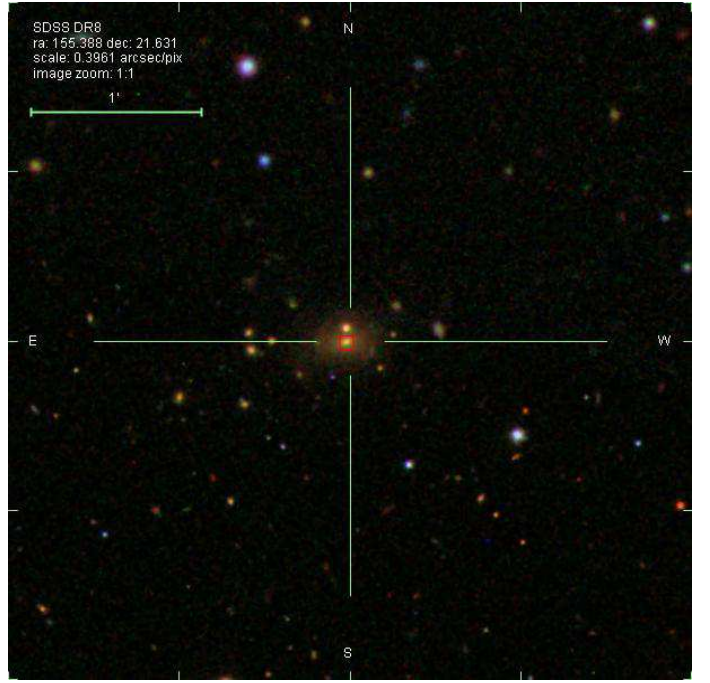}}
  \caption{SDSS colour image of the 2XMM J102133.2+213752 with a field of 
view of four arcmin a side centred on the X-ray peak position 
as indicated by the cross hair. The BCG with a spectroscopic redshift is marked 
by a square and is coincident with the X-ray position.} 
  \label{f:108143_color}
\end{figure}

The main idea of the finding algorithm is to identify the likely brightest cluster 
galaxy (BCG) among the galaxies with similar redshift within one arcmin from 
the X-ray centroid position and then search for an overdensity of surrounding 
member candidates. To confirm the X-ray cluster candidates optically and to 
measure their redshifts, we followed these steps:

\begin{enumerate}

\item We plotted the photometric redshift histogram of all galaxies within one arcmin 
from the X-ray position with $ m_{r} \leq 22$ mag, $ \bigtriangleup m_{r} < 0.5 $ mag  
and the fractional error of the photometric redshift  
$ \bigtriangleup z_{p} / z_{p} < 0.5 $, as shown in Figure~\ref{f:108143_hist1}.

\item We computed a tentative photometric redshift of the cluster as the centre
of the redshift bin in the main peak,  $ z_{p,\,M} $.
To ensure that the distributions of the photometric redshifts of background 
galaxies did not produce this peak in the histogram, we selected 360 random 
positions in the SDSS sky coverage and counted the galaxies with the same  
magnitude and photometric redshift criteria as were used in the previous step 
within one arcmin from the field positions. We chose this large number 
of fields to obtain the average redshift distribution of background 
galaxies.
Figure~\ref{f:360_fields} shows the average distribution of the galaxy counts 
within these fields as a function of redshift. The distribution does not 
exceed 0.91 per redshift bin. It is unlikely that the background 
galaxies have a significant influence on the redshift determination.    
Therefore, we can neglect subtracting the background galaxies in the current 
step to compute a tentative cluster redshift.

\item We identified the BCG as the brightest galaxy of the galaxies within 
one arcmin around the X-ray position with a photometric redshift in the
interval  $ z_{p,\,M} \pm 0.04(1+z_{p,\,M}) $. When the algorithm found 
multiple peaks in the redshift histogram, we selected the BCG candidate 
closest to the X-ray position.
\citet{Wen09} have shown that a redshift interval of $\pm 0.04(1+z_{p,\,M})$
comprises 80 percent of the clusters members. We assumed that our tentative redshift 
gives a less reliable but still robust estimate of cluster membership.
The redshift of the likely BCG does not 
necessarily lie in the peak bin of the redshift histogram, but may be within  
one of the adjacent bins. Therefore, we initially allowed that the BCG
candidate lies either in the central or in one of the adjacent redshift bins.
We then chose as BCG the brightest galaxy in the bins nearest to the X-ray
position.

\item To detect an overdensity of galaxies in 3D space, all galaxies within  
a radius of 560 kpc from the X-ray emission peak within the redshift interval 
$ z_{p,\,BCG} \pm 0.04(1+z_{p,\,BCG})$ were considered as cluster member 
candidates, N($<$560 kpc).  
The redshift range used here is the same as that used by \citet{Wen09}.
Since the physical size of the cluster is not a priori known, 
we chose a radius of 560 kpc as the average of $R_{500}$ from paper I. 
This radius is similar to the radius used by \citet{Wen09} for detecting 
galaxy overdensity. These authors showed using  Monte Carlo simulation tests 
that a radius of 500 kpc gives a high overdensity level and a low false 
detection rate. Because we are not computing the cluster richness in the 
current work, we did not subtract the background galaxies. The identified 
cluster member candidates were only used to compute the cluster redshift.

\item The cluster photometric redshift, $\bar{z}_p$, was finally determined as 
the weighted average of the photometric redshift of N($<$560 kpc) with weights 
given as $w_i = 1/(\bigtriangleup z_{p,\,i})^2$. The redshift value for our 
example cluster is marked by the  vertical red line in 
Figure~\ref{f:108143_hist1}.
When there were available spectroscopic redshifts of N($<$560 kpc), the cluster 
spectroscopic redshift, $\bar{z}_s$, was the weighted average of 
the available spectroscopic redshifts, as indicated by the blue line in 
Figure~\ref{f:108143_hist1}. For the example cluster, only the BCG has 
a spectroscopic redshift. 
Figure~\ref{f:108143_dist} shows the sky distribution of the cluster member 
candidates within 560 kpc from the X-ray centroid; they are represented by 
red dots, and the field galaxies are represented by blue dots.

\item A cluster was considered detected when there were at least eight cluster 
member galaxies within 560 kpc and two members within one arcmin. When 
N($<$560 kpc) $<8$ but the estimated redshift was consistent with either an
available redshift from the literature or spectroscopic redshift from the
current algorithm, we also considered it detected cluster. 
The final decision to confirm the optical cluster detection was made by  
visual inspection of the SDSS colour image of the cluster field, which led  
to the exclusion of misidentified optical counterparts in a few cases. 
Figure~\ref{f:108143_color} shows the SDSS colour image of the example 
cluster with a field of view 4$'$ $\times$ 4$'$ centred at the X-ray position.

\end{enumerate}

\begin{figure}
  \resizebox{\hsize}{!}{\includegraphics{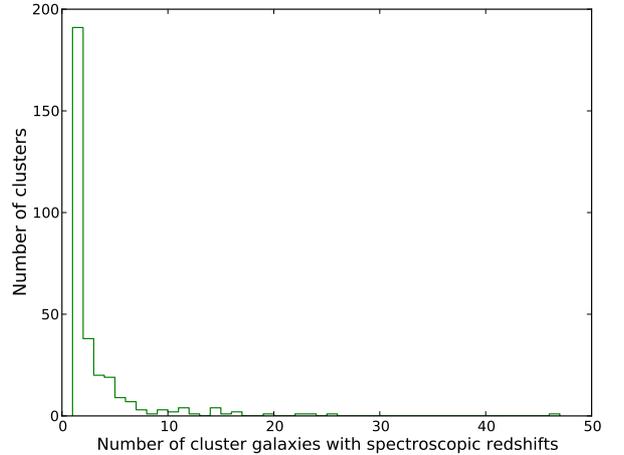}}
  \caption{Distribution of the number of cluster members with spectra of the 
spectroscopically confirmed clusters. The bin size of the histogram is one.} 
  \label{f:Hist_Nzs}
\end{figure}

\begin{figure}
  \resizebox{\hsize}{!}{\includegraphics{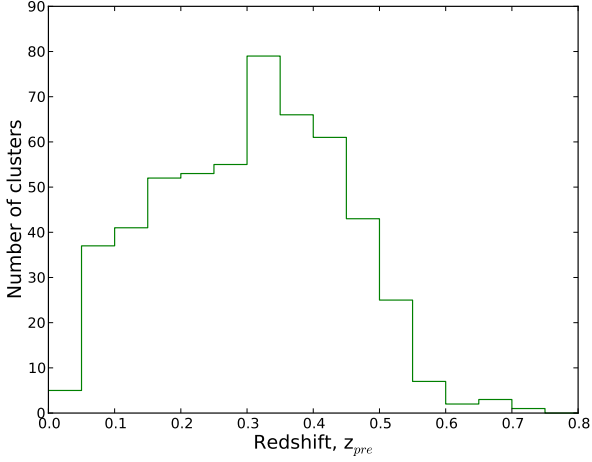}}
  \caption{Distribution of estimated photometric redshifts and, 
where available, the spectroscopic redshifts of the optically confirmed 
cluster sample.}
  \label{f:Hist_zc}
\end{figure}

\begin{figure}
  \resizebox{\hsize}{!}{\includegraphics{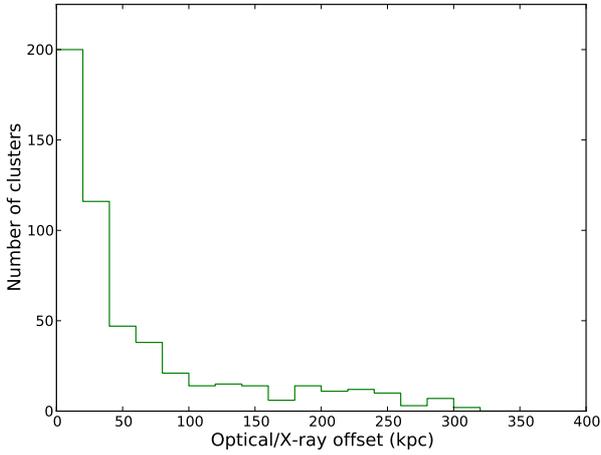}}
  \caption{Distribution of the linear separation between the likely BCG 
and the X-ray emission peak of the optically confirmed cluster sample.} 
  \label{f:Hist_offset}
\end{figure}

Our procedure yielded 530 optically confirmed galaxy clusters with measured
redshifts. We refer to this sample as the optically confirmed cluster sample,
which spans  a wide redshift range from 0.03 to 0.70. About 60 percent of this
sample are spectroscopically confirmed. Figure~\ref{f:Hist_Nzs} shows the
distribution of the number of cluster galaxies per cluster with spectroscopic
redshifts.  
Figure~\ref{f:Hist_zc} shows the distribution of the estimated photometric 
redshifts and, where available, spectroscopic redshifts of the optically confirmed 
cluster sample.  
The projected separation between the X-ray centres and the optical 
centres (chosen to be the BCGs positions) of the cluster sample is 
shown in Figure~\ref{f:Hist_offset}. The distribution has a median offset of
29\,kpc, 86 percent of the BCGs are found within 150\,kpc. 
The maximum projected separation between the BCGs and X-ray peaks is about 320
kpc. The reason for the small observed offset lies in the way of the sample
construction, but the offset distribution seems to agree with the corresponding 
distribution derived for the maxBCG survey and ROSAT clusters \citep{Rykoff08}.

\subsubsection{Redshift uncertainty and comparison with published redshifts}

\begin{figure}
  \resizebox{\hsize}{!}{\includegraphics{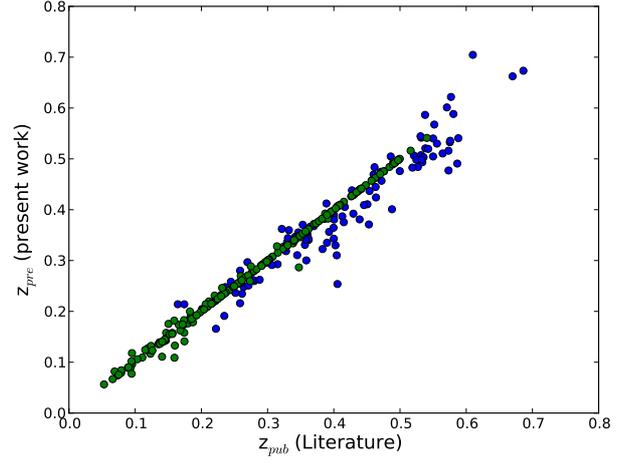}}
  \caption{Comparison between the estimated redshifts, $z_{pre}$,  and 
the published ones, $z_{pub}$, of the optically confirmed cluster sample. 
The green dots represent the clusters with spectroscopic redshifts, while 
blue dots represent the clusters with photometric redshifts only.} 
  \label{f:compz}
\end{figure}

To assess the optical detection algorithm and the estimation of the cluster 
redshift, we queried the NASA Extragalactic Data base (NED) to search for 
published optical redshifts. The NED lists 301 objects including those from 
our paper I. 
Figure~\ref{f:compz} shows the relation between our estimate of 
the redshifts, $z_{pre}$, and the published ones, $z_{pub}$. The clusters 
with available spectroscopic redshifts are represented by the green dots,  
the clusters with photometric redshifts only are represented by 
the blue dots. In general, the newly estimated redshifts agree very well 
with the published ones.

For clusters with a redshift difference $| z_{pre} - z_{pub}| > 0.05$, 
about 5 percent of the sample, we visually re-investigated the colour 
image (as in Figure~\ref{f:108143_color}) and the distribution on the sky 
of the identified cluster members (as in Figure~\ref{f:108143_dist}). 
This led in all cases to a revision of the published redshifts, and we therefore  
regard the newly determined redshifts as more reliable than the published ones, 
which were based on optical search methods alone. We note that the redshifts used 
in paper I also needed to be revised for about 5 percent of the objects for 
the same reason.

\begin{figure}
  \resizebox{\hsize}{!}{\includegraphics{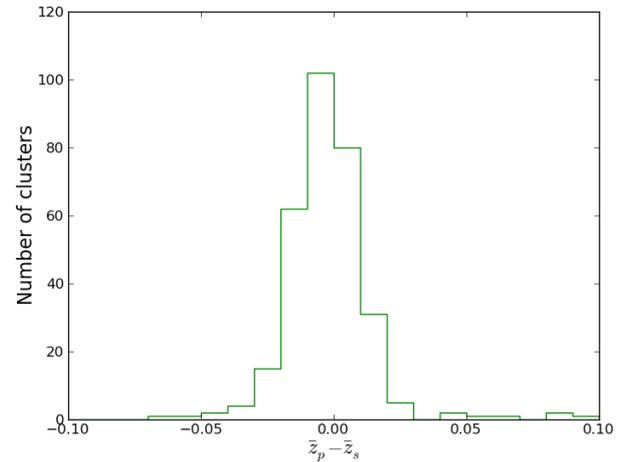}}
  \caption{Distribution of the differences between the photometric, $\bar{z}_p$,
and spectroscopic, $\bar{z}_s$, redshifts of the optically confirmed cluster sample.} 
  \label{f:Hist_zp_zs}
\end{figure}

Of the optically confirmed cluster sample, 310 galaxy clusters are 
spectroscopically confirmed with at least one member galaxy with 
spectroscopic redshift from the existing SDSS data (SDSS-DR8). To assess 
the accuracy of our weighted average photometric redshift, $\bar{z}_p$,
we compared it with the weighted average spectroscopic redshift, $\bar{z}_s$. 
Figure~\ref{f:Hist_zp_zs} shows the distribution of the redshift differences, 
$\bar{z}_p$ - $\bar{z}_s$, of the sample. 
The standard deviation of these redshift differences is 0.02, which roughly  
indicates the accuracy of the estimated photometric redshifts.  
Therefore, we are confident about the reliability of the photometric 
redshift measurements.

\section{X-ray data analysis}

We used a similar procedure to that in paper I to reduce and analyse the X-ray 
data of the optically confirmed cluster sample. The raw XMM-Newton data were 
downloaded using the Archive InterOperability System (AIO), which provides 
access to the XMM-Newton Science Archive \citep[XSA:][]{Arviset02}. 
These data were reprocessed to generate the calibrated and filtered event 
lists for the EPIC (MOS1, MOS2, and PN) cameras with a recent version of the 
XMM-Newton Science Analysis Software (SAS11.0.1). 
To determine the source extraction radii with the highest signal-to-noise
ratio (S/N), we created the radial profiles in the energy band [0.5-2.0] keV
of each  camera as well as for EPIC. Then the S/N was calculated as a function
of radius taking into account the background values as given in the 2XMMi  
catalogue.

The X-ray spectra of each cluster were generated from a region with the 
determined optimum extraction radius, which corresponded to the highest 
EPIC S/N. The spectral extraction from the optimal aperture was chosen 
to reduce the statistical uncertainty in the derived temperatures 
and luminosities from the spectral fits. The background 
spectra were extracted from a circular annulus around the cluster with inner 
and outer radii equaling two and three times the optimum radius, respectively. 
Other field sources embedded in the source and background regions of the 
cluster were removed.

The extracted spectra were binned to one count per bin. Spectra for each 
cluster were simultaneously fitted in XSPEC \citep[version 12.7.0]{Arnaud96} 
with a single-temperature optically thin thermal plasma model modified by 
galactic absorption of neutral matter, $\tt TBabs*MEKAL$ in XSPEC 
terminology \citep{Mewe86,Wilms00}. The temperature and the normalisation 
of the plasma model were allowed to vary while the abundance was fixed at 
0.4 $Z_\odot$. The hydrogen absorbing column density, $nH$,  was 
derived from the Leiden/Argentine/Bonn (LAB) survey \citep{Kalberla05} 
and was fixed to this value. The spectral fit was performed using the Cash 
statistics with one count per bin, a recommended strategy for sources with 
low photon counts \citep[e.g.][]{Krumpe08}. 

To avoid a conversion of the fit to a local minimum of the fitting
statistics, we ran series of fits stepping the temperature from 0.1
to 15 keV  with a step size = 0.05 using the $\tt{steppar}$ command within
XSPEC. We note that when the model spectrum is interpolated from 
a pre-calculated table, the cluster temperatures in some cases tend to 
converge exactly at the temperature grid points of the model table.  
Therefore, we ran the $\tt MEKAL$ code with the option of calculating 
the model spectrum for each temperature during the fitting and stepping process.

The spectral fitting provided us with the X-ray temperatures, fluxes in 
[0.5-2.0] keV, and luminosities in the rest frame energy band [0.5-2.0] keV 
and their errors. 
The errors of the X-ray temperatures, fluxes, and luminosities represent the 68  
percent confidence range. The bolometric X-ray luminosity over the rest frame 
energy range (0.1 to 50.0) keV was determined from the dummy response matrices 
based on the best-fitting model parameters. The fractional error in the 
bolometric luminosity was assumed to be the same as the fractional 
error of the luminosity in the given energy band. 
To confirm that this assumption is valid, we varied the temperatures by 
$\pm$ 1 $\sigma$ in a few cases. We found the measured band luminosities 
to be within their errors.   

We accepted the X-ray parameters (temperature, flux, luminosity) of a 
cluster if the relative errors of both  the temperature and luminosity were
smaller than 50 percent. A final check was made to ensure that neither the
source nor the background region were affected by detector artefacts and/or
astronomical objects. We also visually screened spectral fits applied to
the data and rejected poor spectral fits. 

Finally, the derived bolometric luminosities were used to estimate the cluster 
luminosities and masses at $R_{500}$ through an iterative method, as
briefly described below and as discussed in more detail in paper I.



\section{Results and discussion}

 We were able to derive reliable X-ray parameters from spectral fits for 345 systems 
of the optically confirmed cluster sample. In the next subsections, we compare 
our new results with the common clusters from (a) the XMM Cluster Survey 
\citep{Mehrtens12}, (b) the MCXC catalogue \citep{Piffaretti11}, and 
(c) the paper I sample.  
We then proceed to derive an updated \ltr relation based on this new sample.
For the remaining 185 clusters of the optically confirmed sample
without proper spectral fits, we used the X-ray flux as given in the 2XMMi-DR3
catalogue to estimate the luminosity and mass. We finally present the 
X-ray luminosity-redshift distribution of the whole optically confirmed 
cluster sample.



\subsection{Cluster sample with reliable X-ray parameters from the 
spectral fits}

\begin{figure}
  \resizebox{\hsize}{!}{\includegraphics{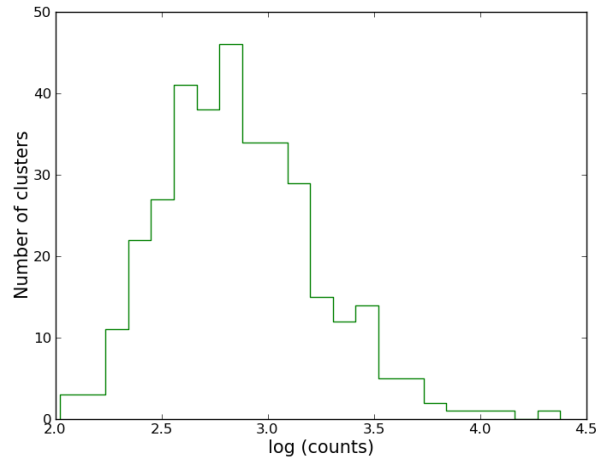}}
  \caption{Distribution of the aperture net EPIC photon counts in 
[0.5-2.0] keV derived from the spectral fitting for the cluster sample 
with X-ray spectroscopic parameters.}
  \label{f:counts-his}
\end{figure}

\begin{figure}
  \resizebox{\hsize}{!}{\includegraphics{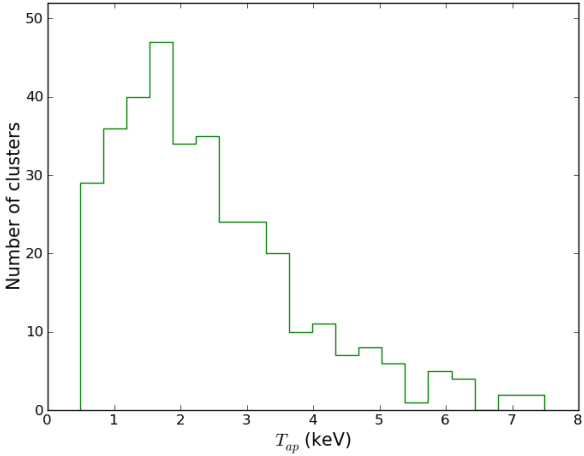}}
  \caption{X-ray spectroscopic temperature distribution of the cluster sample
with reliable X-ray parameters from the spectral fit.} 
  \label{f:kT-his}
\end{figure}

Figure~\ref{f:counts-his} shows the distribution of the net EPIC photon counts 
in the energy interval [0.5-2.0] keV for clusters that could be fitted successfully.
It shows that 87 percent of our
clusters have more than 300 source photons. In some cases a successful
spectral fit could be achieved with just a few more than 100 photons by 
combining clean X-ray data and previous knowledge of the cluster
redshift. Our new sample has a wide range of temperatures from 0.5 to 7.5 keV,
which is shown in Figure~\ref{f:kT-his}. The average relative errors of the 
temperatures and luminosities are 0.20 and 0.06, respectively.

We followed an iterative method (see paper I) to compute the physical parameters 
for each cluster. The estimated aperture X-ray bolometric luminosity and its
error,  optimal extraction radius, and the redshift were used as input 
to determine the radius $R_{500}$, the X-ray bolometric 
luminosity within $R_{500}$, $L_{500}$, and the cluster mass at $R_{500}$,
$M_{500}$.
The main idea of the iterative method is to extrapolate the aperture bolometric 
flux to the bolometric flux at $R_{500}$ based on a $\beta$ model of the form 
\begin{equation}
 S(r) = S(0)\,\Bigg[1 + \Big(\frac{r}{r_{\rm c}} \Big)^2 \Bigg]^{-3\beta + 1/2},
\end{equation}
where $\beta$ and core radius, r$_{\rm c}$, depend on temperature (see Eq.~4 
and 5 in paper I). The correction factor of the flux is used to extrapolate 
the aperture bolometric luminosity to the bolometric $L_{500}$. Finally, 
$M_{500}$ is computed based on the $L_{500}-M_{500}$ relation from 
\citet{Pratt09}. 
The error budget of the estimated $L_{500}$ and $M_{500}$ includes the 
errors of the input parameters, the intrinsic scatter in the \ltr and 
$L_{\rm X}-M$ relations, and the propagated errors of their slopes  and the
intercepts. The median correction factor between the extrapolated bolometric
luminosity to  $R_{500}$ and the aperture bolometric luminosity,
$L_{500}/L_{bol}$, was 1.7.

Table~\ref{tbl:SF_sample}, available in full form at the CDS, represents the 
first ten entries of the X-ray-selected cluster sample with a total of 345 rows.
For each cluster the catalogue lists the cluster 
identification number (detection id, detid) and its name (IAUNAME) in 
(cols.~[1] and [2]), the right ascension and declination of X-ray emission in 
equinox J2000.0 (cols.~[3] and [4]), the XMM-Newton observation id (obsid) 
(col.~[5]), the optical redshift (col.~[6]), the scale at the cluster redshift 
in kpc/$''$ (col.~[7]), 
the aperture and $R_{500}$ radii in kpc (cols.~[8] and [9]), the cluster aperture  
X-ray temperature $T_{ap}$ and its positive and negative errors in keV 
(cols.~[10], [11] and [12], respectively), the aperture X-ray flux 
$F_{ap}$ [0.5-2.0] keV and its positive and negative errors in units of 
$10^{-14}$\ erg\ cm$^{-2}$\ s$^{-1}$ (cols.~[13], [14] and [15], respectively), 
the aperture X-ray luminosity $L_{ap}$ [0.5-2.0] keV and its positive and 
negative errors in units of $10^{42}$\ erg\ s$^{-1}$ (cols.~[16], [17] and 
[18], respectively), the cluster bolometric luminosity $L_{500}$ and its error 
in units of $10^{42}$\ erg\ s$^{-1}$ (cols.~[19] and [20]), the cluster mass 
$M_{500}$ and its error in units of $10^{13}$\ M$_\odot$ (cols.~[21] and [22]), 
the Galactic HI column in units $10^{22}$\ cm$^{-2}$ (col.~[23]), the 
{\tt objid} of the likely BCG in the SDSS-DR8  (col.~[24]), the BCG right 
ascension and declination in equinox J2000.0 (cols.~[25] and [26]), the 
estimated photometric and, where available, the spectroscopic redshift of the cluster 
(col.~[27] and col.~[28]), the number of cluster members within 560 kpc with
available spectroscopic redshifts that were used to compute the cluster
spectroscopic redshift, (col.~[29]), the redshift  type (col.~[30]), the
linear offset between the cluster X-ray position and the BCG position
(col.~[31]), and the NED name and its references (col.~[32] and  col.~[33]).



\subsection{Cluster sample with X-ray flux from the 2XMMi-DR3 catalogue}

For clusters with insufficient X-ray data to perform a proper spectral fit, we  
estimated the X-ray parameters based on the EPIC flux and its error in 
0.5-2.0 keV as given in the 2XMMi-DR3 catalogue. 
The catalogue provides aperture-corrected fluxes that were   
calculated by the SAS tasks {\tt emldetect}. For the individual cameras, 
individual-band fluxes were calculated from the respective band count rate 
using the filter- and camera-dependent energy conversion factors corrected 
for the dead time from the read-out phase. The EPIC flux in each band was
estimated as the mean of the band-specific detections in all cameras weighted 
by their errors \citep{Watson09}. Here we used the combined EPIC flux in band 2 
(0.5 - 1.0 keV) and band 3 (1.0 - 2.0 keV) and its propagated error, 
$F_{\rm cat}$ in [0.5-2.0] keV.

\begin{figure}
  \resizebox{\hsize}{!}{\includegraphics{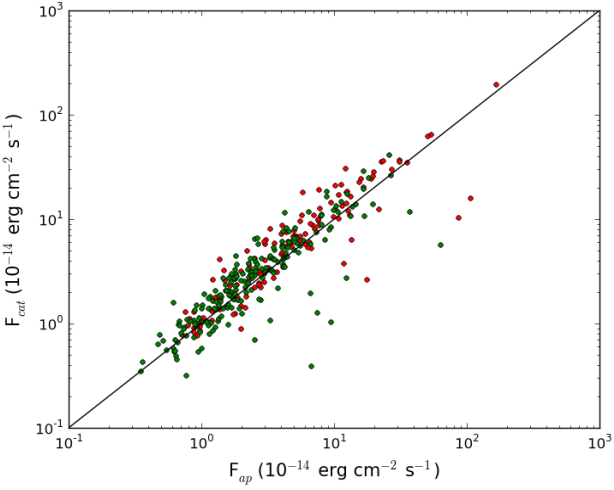}}
  \caption{Cluster flux, $F_{cat}$, in 0.5-2.0 keV from the 2XMMI-DR3 
catalogue plotted against the cluster flux, $F_{ap}$, in 0.5-2.0 keV from 
the best-fitting model parameters for the cluster sample with spectroscopic 
parameters. The red dots represent the first cluster sample from paper I, while 
the green dots represent the extended sample with reliable X-ray parameters.} 
  \label{f:Fxcat-Fxsf}
\end{figure}

Figure~\ref{f:Fxcat-Fxsf} shows the relation between the flux given in the 
2XMMi-DR3 catalogue and the aperture flux determined by us for the 
345 clusters with reliable X-ray parameters from the spectral fits. 
It shows a linear relation between the two flux measurements except for  
some outliers (about 5 percent), which we found to be contaminated by point 
sources in the 2XMMi-DR3 catalogue. In general terms, the catalogued flux 
is higher than the aperture flux because the former was computed for the 
integrated $\beta$-model.

\begin{figure}
  \resizebox{\hsize}{!}{\includegraphics{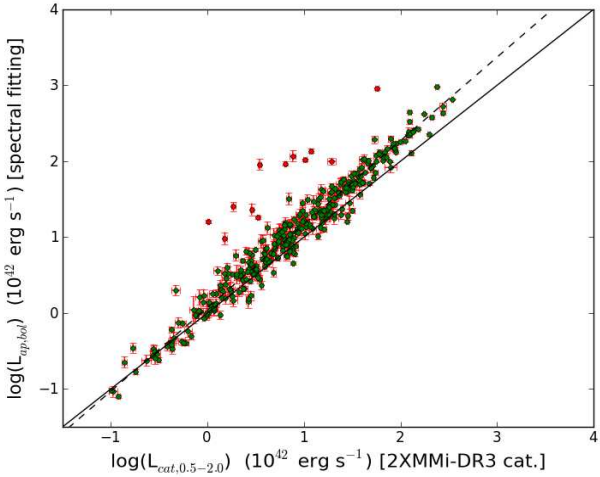}}
  \caption{Aperture bolometric luminosities, $L_{\rm ap,\,bol}$, plotted 
against the 2XMMi-DR3 catalogue luminosities, $L_{\rm cat,\,0.5-2.0}$, 
of the cluster sample with reliable parameters from the spectral fits. 
The one-to-one relationship is represented by the solid line. The dashed line 
represents the best-fit using the BCES orthogonal regression method after 
excluding 12 outliers that are represented by red dots.} 
  \label{f:Lcatband-Lapbol}
\end{figure}

Figure~\ref{f:Lcatband-Lapbol} shows the relation between the aperture 
bolometric luminosities, $L_{\rm ap,\,bol}$, and $L_{\rm cat,\,0.5-2.0}$
of the cluster sample with X-ray spectroscopic parameters, where 
$L_{\rm cat,\,0.5-2.0}$ is based on $F_{\rm cat}$ in [0.5-2.0] keV.
 Generally, there is a linear 
relation between the two luminosities except for 12 outliers with 
$L_{\rm ap,\,bol} / L_{\rm cat,\,0.5-2.0} > 2$.     
Ignoring these outliers, we performed a linear regression between their logarithms 
to convert $L_{\rm cat,\,0.5-2.0}$ to $L_{\rm ap,\,bol}$ for the 185 clusters 
without proper spectral fit. 
The best-fit linear relation derived using the BCES orthogonal regression 
method \citep{Akritas96} is represented by the dashed line in 
Figure~\ref{f:Lcatband-Lapbol} and is given by Eq.~2, 
\begin{equation}
  \log\ (L_{\rm ap,\,bol}) = 0.07 + 1.10\ \log\ (L_{\rm cat,\,0.5-2.0}).  
\end{equation}

\begin{figure}
 \resizebox{\hsize}{!}{\includegraphics{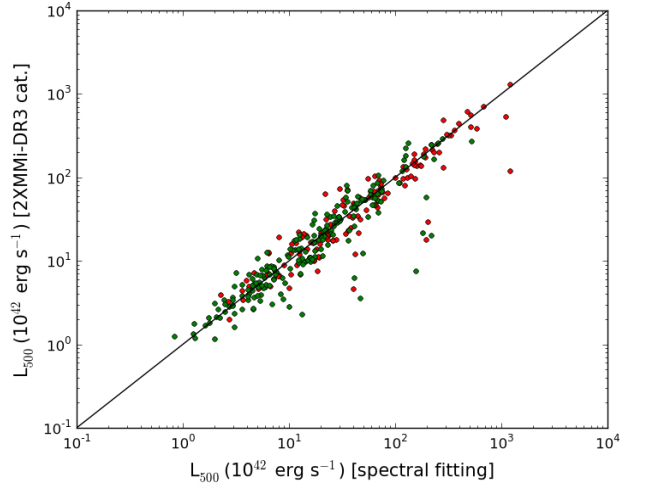}}
 \caption{Comparison between the measured bolometric luminosity $L_{500}$ based 
on the flux given in the 2XMMi-DR3 catalogue and the bolometric $L_{500}$ based 
on the spectral fit flux for the first (red dots) and extended (green dots) 
cluster sample with X-ray spectroscopic parameters from the survey. The solid 
line shows the one-to-one relationship.} 
 \label{f:L500cat-L500sf}
\end{figure}

Using the iterative method described above, we computed the bolometric $L_{500}$ 
per cluster with the redshift, aperture radius $R_{ap}$ and aperture bolometric 
luminosity $L_{\rm ap,\,bol}$ as input. The aperture radius used here is still 
the radius that corresponds to the highest EPIC S/N, see Section 3. 
We finally determined for each cluster $R_{500}$, $M_{500}$, and $T_{500}$ 
and the corresponding errors using the extrapolated values for $L_{500}$. 
From the comparison between the bolometric $L_{500}$ based on the catalogue 
flux and the bolometric $L_{500}$ based on the spectroscopic flux, there is 
no obvious systematic difference between the two luminosities, as shown in 
Figure~\ref{f:L500cat-L500sf}. Therefore, the conversion from 
$L_{\rm cat,\,0.5-2.0}$ to $L_{\rm ap,\,bol}$ and the iterative 
procedure are acceptable.

Table~\ref{tbl:cat_sample}, a full version of which is provided at the CDS,  
represents the first 10 entries of the X-ray-selected cluster sample 
with X-ray parameters based on the given flux in the 2XMMi-DR3 
catalogue, comprising 185 clusters. For each cluster, the catalogue 
provides the cluster identification number (detection id, detid) and its name 
(IAUNAME) in (cols.~[1] and [2]), the right ascension and declination of 
X-ray emission in equinox J2000.0 (cols.~[3] and [4]), the XMM-Newton 
observation id (obsid) (col.~[5]), the optical redshift (col.~[6]), the scale 
at the cluster redshift in kpc/$''$ (col.~[7]), the $R_{500}$ in kpc 
(col.~[8]), the 2XMMi-DR3 X-ray flux $F_{cat}$ [0.5-2.0] keV and its error in 
units of $10^{-14}$\ erg\ cm$^{-2}$\ s$^{-1}$ (cols.~[9], and [10]), 
the estimated X-ray luminosity $L_{cat}$ [0.5-2.0] keV and its error 
in units of $10^{42}$\ erg\ s$^{-1}$ (cols.~[11], and [12]), 
the cluster bolometric luminosity $L_{500}$ and its error 
in units of $10^{42}$\ erg\ s$^{-1}$ (cols.~[13] and [14]), the cluster mass 
$M_{500}$ and its error in units of $10^{13}$\ M$_\odot$ (cols.~[15] and [16]), 
the $T_{500}$ and its error in units of keV (cols.~[17] and [18]),
the {\tt objid} of the likely BCG in SDSS-DR8 (col.~[19]), the BCG right 
ascension and declination in equinox J2000.0 (cols.~[20] and [21]), the 
estimated photometric and, where available,  the spectroscopic redshift of the 
cluster (col.~[22] and col.~[23]), the number of 
cluster members within 560 kpc with available spectroscopic redshifts that were
used to compute the cluster spectroscopic redshift, (col.~[24]), the redshift 
type (col.~[25]), the linear offset between the cluster X-ray position and the 
BCG position (col.~[26]), and the NED name and its references (col.~[27] and 
col.~[28]).

\subsection{Analysis of the cluster sample with reliable X-ray parameters }

We present a comparison of the measured parameters (temperatures, luminosities, 
and masses) of the cluster sample (345 systems) that have reliable X-ray 
parameters from the spectral fits with the values available in the literature.

\subsubsection{Comparison with the XCS sample}

\begin{figure}
  \resizebox{\hsize}{!}{\includegraphics{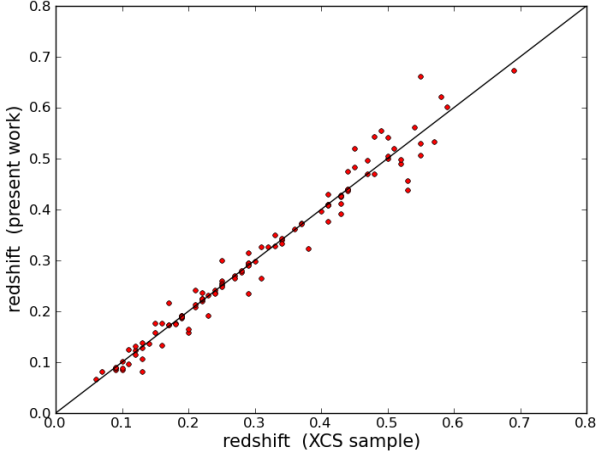}}
  \caption{Comparison of the estimated redshifts of the common sample 
between the XCS catalogue and our sample with X-ray spectroscopic parameters. 
The solid line in the figure indicates the unity line.} 
  \label{f:zs-zxcs}
\end{figure}

The largest published catalogue of X-ray clusters so far, based on the entire 
XMM-Newton archive, was compiled by the XMM Cluster Survey team
\citep[XCS,][]{Romer01, Lloyd-Davies11, Mehrtens12}. The catalogue consists 
of 503 optically confirmed clusters. Of these, 463 systems have redshifts in 
the range 0.05 to 1.46. The X-ray temperatures were measured for 401 clusters.
We cross-matched our cluster sample and the XCS sample with available 
temperature measurements within a matching 
radius of 30 arcsec, which yielded 114 common clusters. About half of the 
common sample was previously published by us in paper I. The standard 
deviation of the redshift difference $(z_{\rm XCS} - z_{\rm pre})$ is 0.027 
and is thus of about the photometric redshift accuracy. There is no systematic 
deviation for instance as a function of redshift, as shown in 
Figure~\ref{f:zs-zxcs}.

\begin{figure}
  \resizebox{\hsize}{!}{\includegraphics{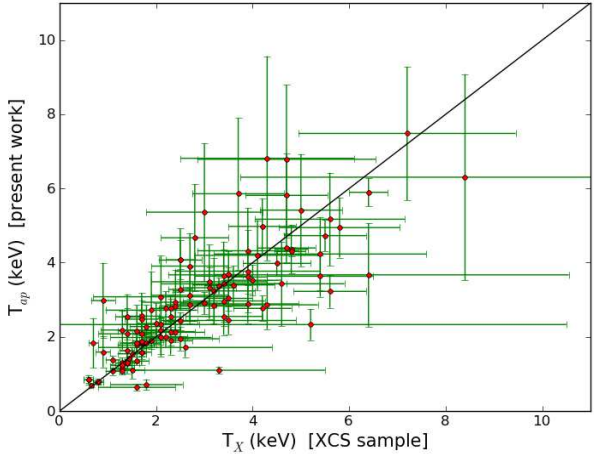}}
  \caption{Comparison of measured temperatures between $T_{ap}$ in our sample 
and $T_{X}$ in XSC sample. The solid line shows the one-to-one relationship. 
The errors are the average errors of positive and negative errors, which are 
provided by the spectral-fitting.} 
  \label{f:Ts-Txcs}
\end{figure}

Even though we extracted the cluster spectra from a different aperture than 
the aperture used in the XCS project and used a different spectral fitting 
procedure, the temperature measurement in general agrees.  
Figure~\ref{f:Ts-Txcs} shows the comparison of
the measured temperatures from  the two projects. We plot the symmetric errors
of each temperature as the average of the positive and negative errors. Our
procedure reveals a slightly smaller temperature uncertainty than derived in
the XCS, with a median 
$\bigtriangleup T_{\rm pre}/\bigtriangleup T_{\rm XCS} = 0.84$. 
The differences between the two temperature measurements have a mean of 
0.02 keV and a standard deviation of 0.93 keV that is similar to the 
standard deviation, 0.82 keV, of the error measurements in temperatures of 
the XCS sample.

\begin{figure}
  \resizebox{\hsize}{!}{\includegraphics{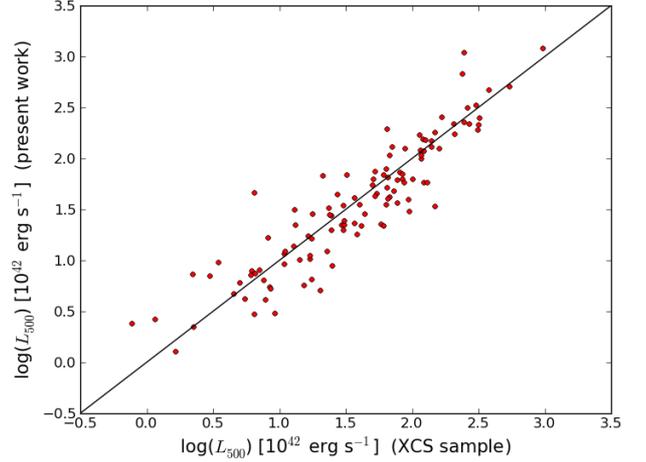}}
  \caption{Comparison between the bolometric luminosities, $L_{500}$, from the 
present work and the corresponding ones from the XCS sample. The solid line 
shows the one-to-one relationship. } 
  \label{f:Ls-Lxcs}
\end{figure}

In the XCS project, the cluster luminosity $L_{500}$ was calculated by using an 
analytical function of the $\beta$ model fitted to the surface brightness profile. 
The same profile was used to determine a scaling factor between the aperture 
luminosity and $L_{500}$ \citep{ Lloyd-Davies11}. Our procedure of the 
extrapolation is described above and is discussed in more detail in
paper I. We found a good agreement between both determinations of $L_{500}$, 
as shown in Figure~\ref{f:Ls-Lxcs}. The ratio between the current luminosity 
measurements to that of the XCS has a median of 0.93.

\subsubsection{Comparison with the MCXC sample }

The MCXC catalogue, a meta-catalogue of X-ray-detected galaxy clusters, 
is compiled from published ROSAT All Sky Survey-based and serendipitous cluster 
catalogues \citep{Piffaretti11}. The catalogue comprises 1743 clusters that
span  a wide redshift range up to 1.3. For each cluster the catalogue lists
redshift, luminosity $L_{500}$ in the 0.1-2.4 keV band, total mass $M_{500}$,  
and radius $R_{500}$. 
Within a cross-matching radius of the cluster centres of 30 arcsec there are
only 23 common clusters. The small overlap is mainly due to our small survey
area and our strategy of investigating serendipitous clusters only, not cluster
targets.

\begin{figure}
  \resizebox{\hsize}{!}{\includegraphics{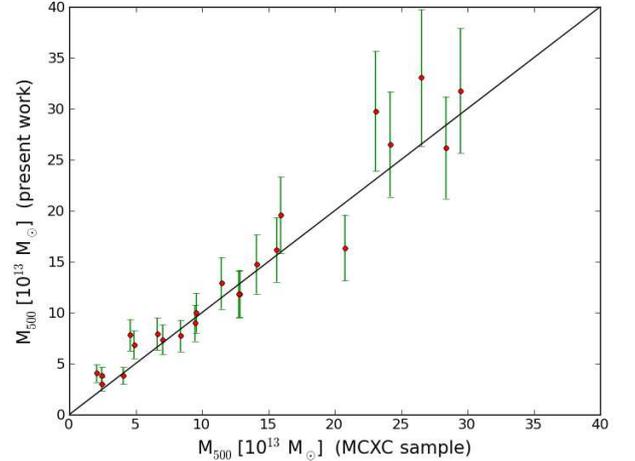}}
  \caption{Comparison of our sample mass estimates within $R_{500}$ with the 
estimated values from the MCXC catalogue. The solid line represents the 
one-to-one relationship.} 
  \label{f:Ms-Mmcxc}
\end{figure}

We compare the masses of the common sample in  Figure~\ref{f:Ms-Mmcxc} and
find consisting results. This comparison makes sure that our mass measurements 
are reliable and not affected by any systematic bias. We also find consistency 
between the redshifts used in both catalogues.

\subsubsection{Comparison with the sample from paper I}

Because we developed an algorithm for estimating the redshifts of the X-ray 
cluster candidates, the redshifts of the first cluster sample from the 
survey, paper I, were revised, as discussed in Section 2. We also revised 
the X-ray spectroscopic parameters for the first cluster sample. The sample in 
common between the current sample with reliable X-ray parameters and the first 
cluster sample consists of 141 systems. The remaining 34 clusters from paper I 
did not pass the quality criterion applied in the present work. 
These missed clusters are nevertheless included in the published cluster
catalogue from this paper with less reliable parameters (see above).

\begin{figure}
  \resizebox{\hsize}{!}{\includegraphics{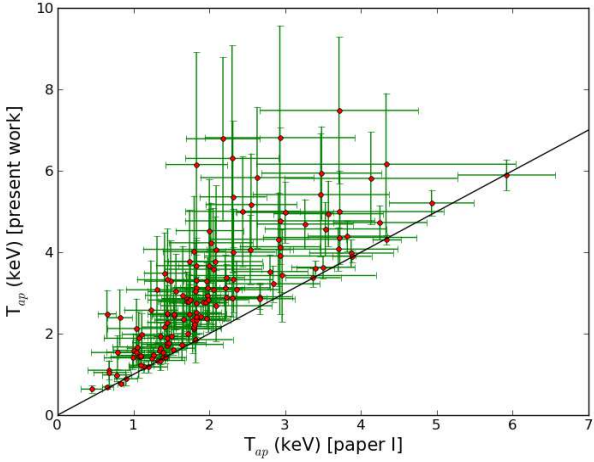}}
  \caption{Current X-ray aperture temperature estimates plotted 
against the corresponding ones of the first cluster sample from paper I. 
To facilitate the comparison, we plot the unity line.} 
  \label{f:kT1-kT2}
\end{figure}

We found a systematic bias of the temperature measurements of the sample in
paper I and the current sample (Figure~\ref{f:kT1-kT2}). When investigating 
possible reasons for the discrepancy we realized that the X-ray data in Paper
I were analysed in an inappropriate manner. The X-ray spectra were not 
grouped and  binned before we applied a spectral model. This led to 
a systematic shift towards too low values of many of the derived temperatures 
determined in paper I.

\begin{figure}
  \resizebox{\hsize}{!}{\includegraphics{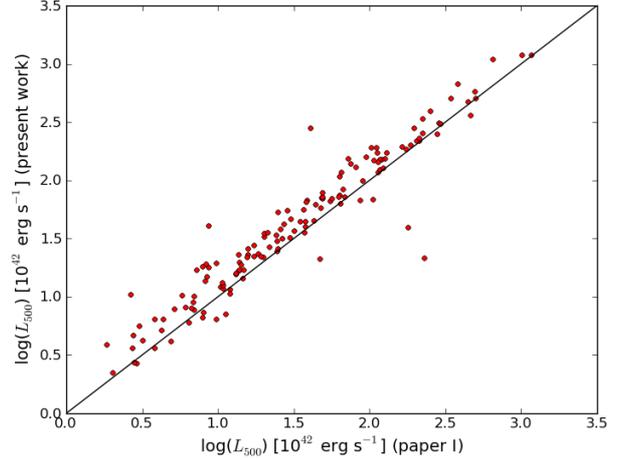}}
  \caption{Comparison between the bolometric luminosities, $L_{500}$, from the 
present work and the corresponding ones from paper I. The solid line shows the 
one-to-one relationship.} 
  \label{f:Lp1-Lp2}
\end{figure}

As a consequence, the luminosities were biased towards lower values
(Figure~\ref{f:Lp1-Lp2}) by a factor of 20 percent. Revised redshifts 
and in some cases revised spectral extraction regions led to a few 
outliers in that figure.

We also presented in paper I the \ltr of the first cluster sample, which we
now regard as affected by the underestimated X-ray temperatures. We are
confident through several sanity checks that our updated temperatures and 
luminosities are reliable, and we re-determine the \ltr relation based on 
the much enlarged and more reliable sample in the next subsection.

\subsection{\ltr relation of the cluster sample with reliable X-ray 
parameters}

Based on the cluster sample with X-ray spectroscopic parameters, we investigate 
the \ltr relation as well as the evolution of its slope and intrinsic scatter 
as presented in the following subsections.

\subsubsection{\ltr relation of the full sample}

The bolometric luminosities $L_{500}$ and the aperture temperatures $T_{ap}$ based on 
X-ray spectral fits were used to investigate the $L_{500}-T_{ap}$ relation 
for the cluster sample with reliable X-ray parameters. 
We note that we were unable to attempt to excise the cores in most cases 
of the cluster sample because of the low resolution of the X-ray 
optics of the XMM-Newton telescopes, short exposure times, and the 
very large distance of most of our clusters. This caveat needs to be 
made when comparing our results with those in the literature, which are 
partly based on nearby clusters with Chandra follow-up.

\begin{figure*}
  \resizebox{\hsize}{!}{\includegraphics{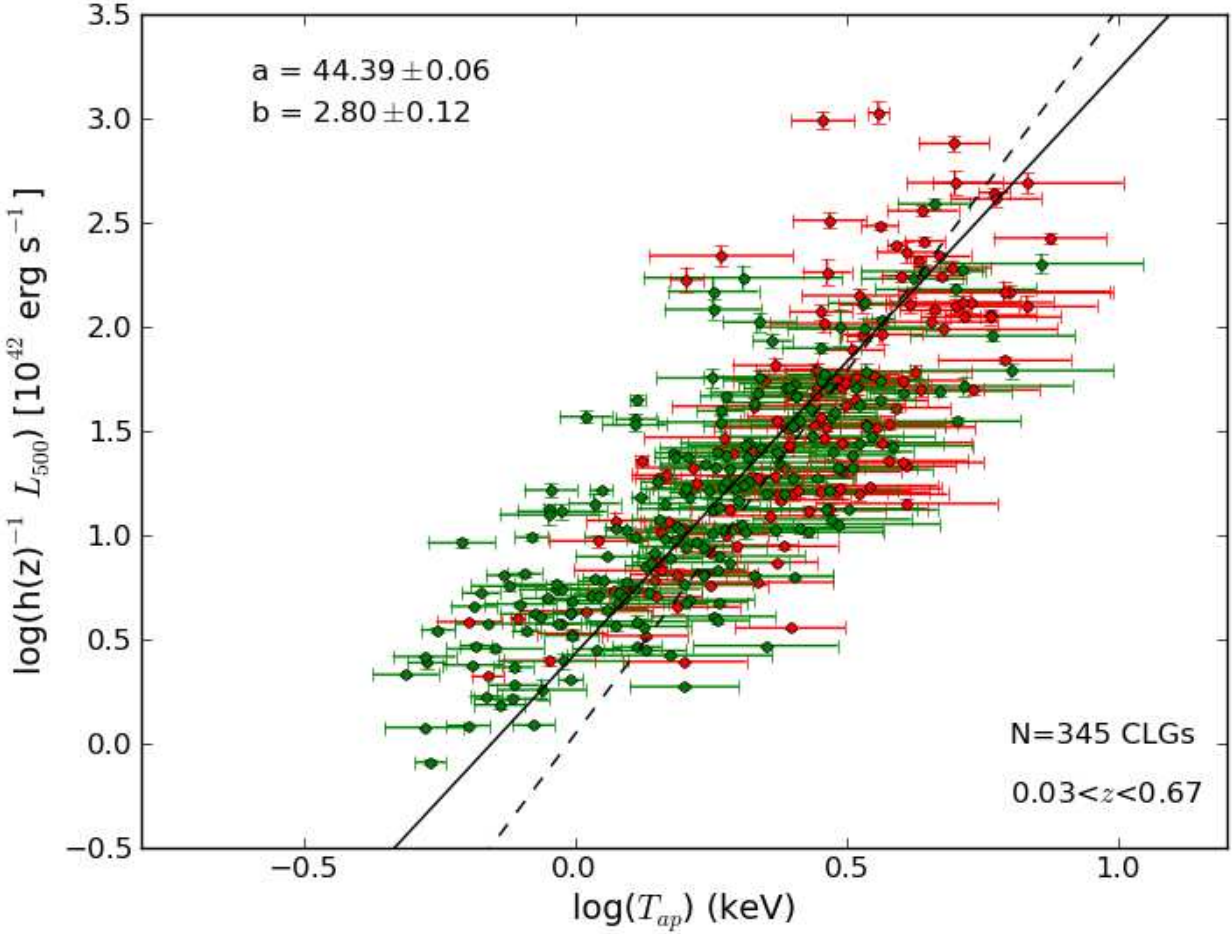}}
  \caption{X-ray bolometric luminosity, $L_{500}$ , plotted against 
aperture X-ray spectroscopic temperature, $T_{ap}$, for the first
(red dots) and expanding (green dots) cluster sample with reliable X-ray parameters. 
The solid line is the fit to both samples using the BCES orthogonal regression.
The intercept, a, and the slope, b, of the fitted line are written in the upper 
left corner of the figure, while the sample number, N, and its redshift range are 
written in the lower right corner. The dashed line represents the relation fit 
of the first cluster sample using the current parameter measurements.}
  \label{f:L-T}
\end{figure*}

Figure~\ref{f:L-T} shows the relation between the measured X-ray bolometric
luminosity, $L_{500}$, modified with the evolution parameter for self-similar
evolution and the X-ray aperture temperature, $T_{ap}$. We used the BCES
orthogonal regression method \citep{Akritas96} to derive the best-fit linear
relation between the logarithms of $L_{500}$ and $T_{ap}$ taking into account
their errors as well as the intrinsic scatter. The best fit is shown in 
Figure~\ref{f:L-T} and is given by Eq.~3,  
\begin{equation}
\log\ (h(z)^{-1}\ L_{500}) = (44.39 \pm 0.06)  + (2.80 \pm 0.12)\ \log\ (T_{ap}/5),
\end{equation} 
where $h(z)$  is the Hubble constant normalised to its present-day value, $
h(z) = \bigl[\Omega_{\rm M} (1+z)^{3} + \Omega_{\Lambda}\Bigr]^{1/2}$, $L_{500}$
in erg s$^{-1}$, and $T_{ap}$ in keV.
By an analysis of objects in common between our list and that of the XCS 
we have shown that our $T_{ap}$ and $L_{500}$ agree well with $T_{\rm X}$ 
and $L_{500}$ of the XCS sample. Not unexpectedly, the slopes and intercepts 
of the corresponding \ltr relations agree with each other within 1-2 $\sigma$ 
(see subsection 4.4.2).

Relations between the luminosity and the temperature, $L_{500}-T_{500}$, 
were published for the REXCESS and HIFLUGCS sample \citep{Pratt09,Mittal11}. 
The REXCESS sample 
comprises 31 nearby ($ z < 0.2 $) galaxy clusters with a temperature range 
from 2 to 9 keV that have been observed with the XMM-Newton. The HIFLUGCS
sample comprises the 64 brightest galaxy clusters in the sky with $kT \ge
1$\,keV and $ z \le 0.2 $, with high-quality {\tt Chandra} data. In both
samples the core emission could be excised but \ltr were published for the
non-excised data as well. The REXCESS team used a fitting procedure similar to
ours, the HIFLUGCS sample was fitted with a BCES-bisector routine.

The present slope of the relation in Eq.~(3), 2.80 $\pm$ 0.12, is slightly
lower than that from the REXCESS sample, 3.35 $\pm$ 0.32, but is still within
1.8$\sigma$. We also found that the present slope agrees with the slope given
by \citet{Mittal11}, $2.94 \pm 0.16$. 

The current cluster sample includes groups with much lower luminosity than
REXCESS and HIFLUGCS, which might influence the slope of the \ltr relation.
If we exclude systems with luminosities $ L_{500} < 5 \times 10^{42}$ erg
s$^{-1}$, the slope of the relation becomes $3.07 \pm 0.19$, which agrees 
much better with the published ones for the REXCESS, HIFLUGCS, and XCS samples.  
The normalisation of the relation, 
$44.46 \pm 0.07$, is still much lower than that of the REXCESS sample, 
$44.85 \pm 0.06$. This is because a much wider temperature range is 
covered by the current large sample, in addition to establishing the current 
relation based on aperture temperatures that are in general slightly higher 
than the temperatures at $R_{500}$. We found that the median scaling factor 
of $T_{ap}$ and $T_{500}$ of the full sample, $T_{ap}/T_{500}$,  was 1.2, where  
$T_{500}$ is the predicted temperature based on the $L_{500}-T_{500}$ 
relation by \cite{Pratt09} using our value for $L_{500}$.

\cite{Eckmiller11} found that the slope of \ltr relation of galaxy groups 
(26 systems, $ L_x \sim 1 - 26 \times 10^{42}$\, erg s$^{-1}$ ) is slightly 
shallower than that derived for clusters (HIFLUGCS), but they are still 
consistent within the errors. These authors found no significant change 
either of the slope derived from a sample that combined groups and clusters 
to a sample consisting of clusters only, which is consistent with the results reported 
by \cite{Osmond04}. We found that the slope derived from the current sample 
(including groups and clusters) agrees well with the slope obtained from 
clusters only (HIFLUGCS sample), but is lower than the slope of the REXCESS 
sample.

The current slope of the \ltr relation is significantly lower than the one 
published in paper I, $3.41\pm 0.15$. The redshifts, temperatures, and 
luminosities of the previous sample were revised. Using the updated values,  
we still find a very steep slope of $3.48 \pm 0.30$, thus confirming
the initial result (the new fit is shown with a dashed line in
Figure~\ref{f:L-T}). 
The much shallower slope found here based on 
the full sample is clearly due to the inclusion of the new objects, which  
have a wider temperature and luminosity range. As discussed above, when 
excluding the low-luminosity systems from the full sample, the slope 
becomes steep, $3.07 \pm 0.19$, which is consistent within 1.4$\sigma$ 
with the updated slope, $3.48 \pm 0.30$, for the paper I-sample.

To estimate the intrinsic scatter in the \ltr relation, we followed the method 
used by \citet{Pratt09}. First we estimated the raw scatter using the 
error-weighted orthogonal distances to the regression line  \citep[Eqs.~3 and
  4 in][]{Pratt09}. Then we computed the intrinsic scatter as the mean value
of the quadratic differences between the raw scatters and the statistical
errors.  The error of the intrinsic scatter was computed as the standard error
of its  value. The computed intrinsic scatter value of the relation, 
$0.48 \pm 0.03$, is slightly higher than the value of REXCESS sample, 
$0.32 \pm 0.06$.

The updated \ltr relation was derived for the first time from  a sample comprising 
345 clusters drawn from a single survey and spans a wide redshift range 
($0.03 < z < 0.67$). Of these, 210 clusters have spectroscopic redshifts for at 
least one cluster member galaxy. The redshifts and X-ray parameters of the 
sample are measured in a consistent way. The sample has X-ray spectroscopic 
temperature measurements from 0.5 to 7.5 keV and a bolometric luminosity range 
of $L_{500} \sim 1.0 \times 10^{42} - 1.0 \times 10^{45}$\ erg s$^{-1}$.

Based on the SDSS, we were able to identify only about half of our X-ray cluster
candidates. The other 50 percent probably represent a more luminous population. 
The omission of that subsample may have an influence on the \ltr relation that 
is yet to be quantified. However, including luminous distant clusters 
does not have a significant effect on the slope of the \ltr relation 
\citep{Hilton12}, as described in the subsequent section.  Moreover, the 
current sample does not include distant clusters beyond $z=0.7$, 
therefore we defer measuring the evolution of the normalisation of 
the \ltr relation to a future study.



\subsubsection{Evolution of the slope and intrinsic scatter}

Based on the first data release of the XCS, \citet{Hilton12} investigated a
possible evolution of the slope and intrinsic scatter of the \ltr relation
in three redshift bins. A sample of 211 clusters with spectroscopic redshift
up to 1.5 was used for this exercise. No evidence for evolution in
either the slope or intrinsic scatter as a function of redshift was found.

Using our much larger sample of clusters with measured X-ray spectroscopic 
parameters, we also investigated a possible evolution of the mentioned 
parameters of the \ltr relation. We divided our sample into three subsamples 
with redshift bins similar to those used by \citet{Hilton12}, 
$0.03 \le z < 0.25, 0.25 \le z < 0.5$, and
$0.5 \le z < 0.7$. The numbers of clusters per redshift bin are listed in
Table~\ref{tbl:LTrelation}. Our two low-redshift subsamples are about twice 
as large as the XCS corresponding subsamples. The number of clusters in the 
high-redshift bin are similar, but the XCS comprises clusters up to 
redshift 1.5. In general, there are about 75 cluster in common between our 
sample and the XCS-DR1 sample that were used to derive the \ltr relation. 
Of these clusters in common, 44 systems were published from our survey in 
paper I.

\begin{figure*}
  \resizebox{\hsize}{!}{\includegraphics[viewport=15  95  570 320, clip]{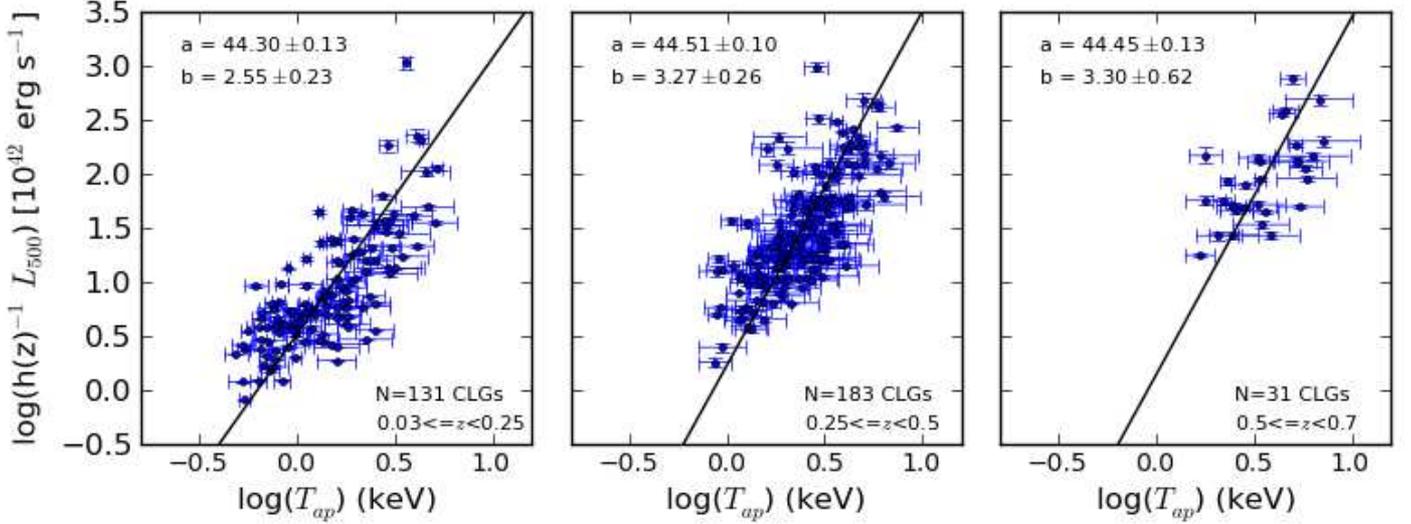}}
  \caption{$L_{500}-T_{ap}$ relations for the three subsamples in redshift bins. 
The redshift bin and the cluster number of these subsamples are written in the 
lower right of the figure. The best-fit line of the subsamples is presented 
by the solid line, while their parameters (intercepts, a, and slopes, b) are 
written in the upper left corner.} 
  \label{f:L-T-zbins}
\end{figure*}

The \ltr relations of our subsamples are shown in Figure~\ref{f:L-T-zbins}. 
When we fitted these susamples using the BCES orthogonal regression method, 
we found that the relation slope of the subsamples in the intermediate- and 
high-redshift bins are consistent, while the subsample in the lowest redshift 
bin has a shallower slope. The reason is that the low-redshift subsample 
includes groups/clusters with low temperature and luminosity, which produces a
shallower slope. Moreover, the present slope of the low-redshift subsample is 
lower than the published one of the corresponding XCS-DR1 subsample and those 
of the REXCESS and HIFLUGCS samples. On the other
hand, the slopes of the intermediate- and high-redshift subsamples agree 
with the slopes of the corresponding XCS-DR1 subsamples.
The intrinsic scatter of all subsamples agree with each other.
Table~\ref{tbl:LTrelation} also lists the fitted parameters (intercept and
slope) of the \ltr relations and their intrinsic scatter together with 
published values (slope, sample size, reference). 

When we fit the \ltr for the low-redshift subsample after excluding the groups 
with low luminosity (i.e. $ L_{500} < 5 \times 10^{42}$ erg s$^{-1}$), the 
slope of the relation becomes $2.86 \pm 0.41$, which agrees with the 
intermediate- and high-redshift subsamples as well as with the corresponding 
published slopes given in Table~\ref{tbl:LTrelation}.
We thus confirm the finding by \cite{Hilton12} that the \ltr relation 
does not show a significant change of its slope and its intrinsic scatter 
as a function of redshift.

\addtocounter{table}{+2}
\begin{table*}
\caption{Fit parameters of the $L_{500}-T_{ap}$ relation, derived from the 
BCES orthogonal regression method, for the three subsamples in redshift bins. 
The fitted model is $\log\ (h(z)^{-1}\ L_{500}) = a  + b\ \log\ (T_{ap}/5)$, 
and the fit parameters (a and b) are also shown in the legend of 
Figure~\ref{f:L-T-zbins}.}

\label{tbl:LTrelation}     
\centering                        
\begin{tabular}{c c c c c c c c c}       
\hline\hline                        
     redshift range & N$_{CLGs}$ & intercept & current slope &  $\sigma_{log}L_{500}$  &published slope &  N$_{CLGs, pub.}$ & ref. \\   
\hline  
 0.03 $\le$ z $<$ 0.25  & 131 &44.30 $\pm$ 0.13 & 2.55 $\pm$ 0.23 & 0.45 $\pm$ 0.04 &  3.18 $\pm$ 0.22 & 96 & 1 \\ 
                        &     &                 &                 &                 &  3.35 $\pm$ 0.32 & 31 & 2 \\ 
                        &     &                 &                 &                 &  2.94 $\pm$ 0.16 & 64 & 3 \\                            
 0.25 $\le$ z $<$ 0.50  & 183 &44.51 $\pm$ 0.10 & 3.27 $\pm$ 0.26 & 0.49 $\pm$ 0.04 &  2.82 $\pm$ 0.25 & 77 & 1 \\ 
 0.50 $\le$ z $<$ 0.70  & 31  &44.45 $\pm$ 0.13 & 3.30 $\pm$ 0.62 & 0.41 $\pm$ 0.07 &  2.89 $\pm$ 0.45 & 38 & 1 \\ 

\hline 
\end{tabular}
\tablebib{1- \cite{Hilton12}; 2- \cite{Pratt09}; 3- \cite{Mittal11}.
}
\end{table*}

\subsection{Distribution of the luminosity with redshift}

\begin{figure*}
  \resizebox{\hsize}{!}{\includegraphics{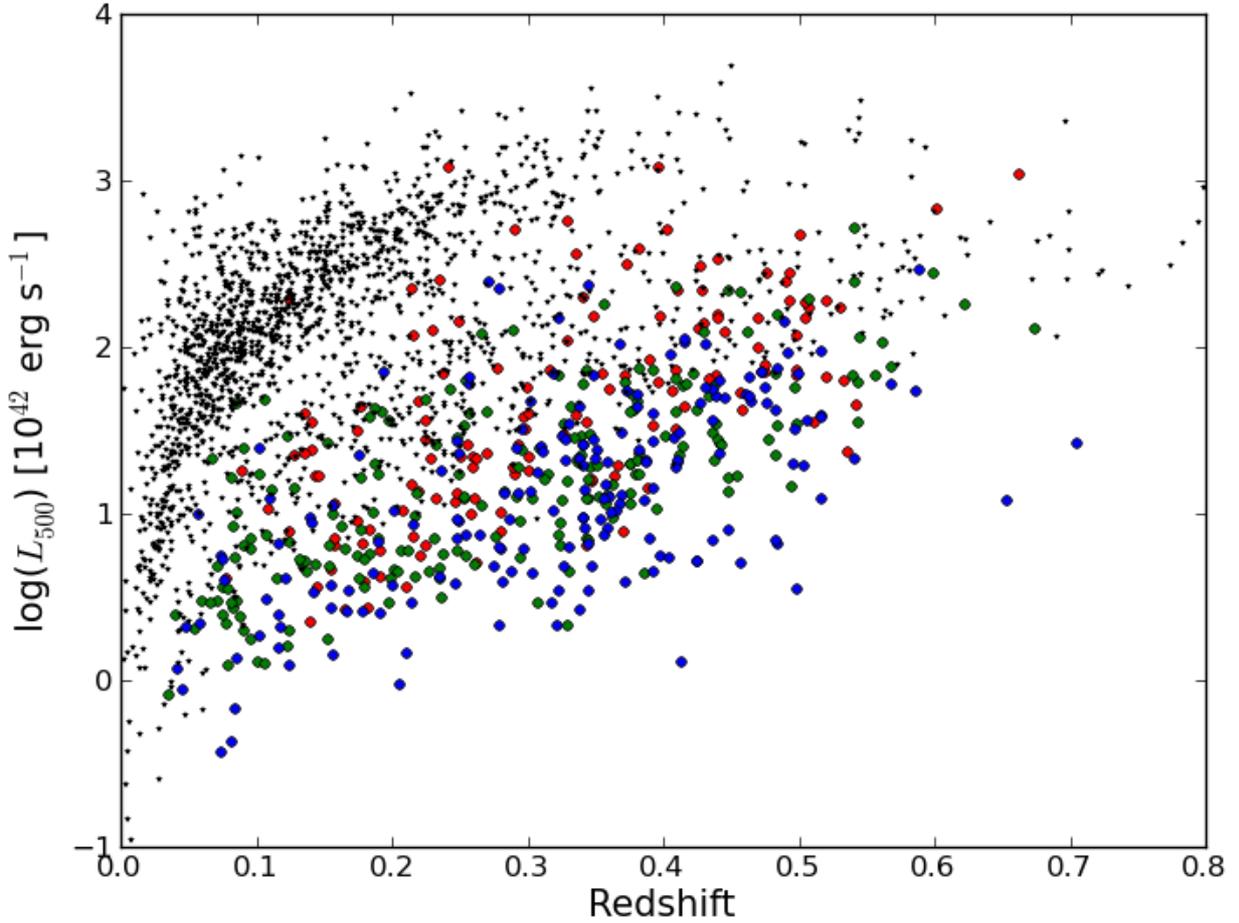}}
  \caption{Distribution of the estimated bolometric luminosity, $L_{500}$,  
as a function of the redshift for the first (red dots) and extended (green dots) 
cluster sample with X-ray spectroscopic parameters, the cluster sample 
(blue dots) with X-ray parameters based on the given flux in the 2XMMi-DR3 
catalogue, and the MCXC cluster sample (black stars) \citep{Piffaretti11}.} 
  \label{f:L-z-all}
\end{figure*}

Figure~\ref{f:L-z-all} shows the distribution of the bolometric luminosity
$L_{500}$ as a function of the redshift for all 530 clusters with redshifts
that were determined in the present work. 
Included are also the 1730 systems from the MCXC catalogue below redshift 0.8.
The X-ray luminosity $L_{500}$ in $0.1-2.4$ keV of the MCXC sample was 
converted into the bolometric luminosity $L_{500}$ by assuming the factor 
$L_{\rm bol,\,500} / L_{\rm band,\,500} = 1.3$. This factor was derived as 
the median of $L_{\rm bol,\,500} / L_{\rm band,\,500}$ for the 23 systems 
in common between the cluster sample with reliable parameters from the 
spectral fitting and MCXC catalogue.

It is obvious that our X-ray-selected samples extend to include 
groups and clusters with low luminosity. The sensitivity of XMM-Newton 
and deeper exposures for some fields allow us to detect less-luminous 
clusters over the redshift range, as shown in Figure~\ref{f:L-z-all}.

\section{Summary and outlook}

We have presented the optically confirmed cluster sample of 530 galaxy 
groups and clusters from the 2XMMi/SDSS Galaxy Cluster Survey. The survey 
consists of 1180 X-ray cluster candidates with at least 80 net photon counts 
selected from the second XMM-Newton serendipitous source catalogue (2XMMi-DR3) 
that are located in the footprint of the SDSS-DR7. The survey area is 210 deg$^2$ 
considering the XMM-Newton field of view has a radius of 15 arcmin.
We developed a finding algorithm for detecting the optical counterparts of 
the X-ray cluster candidates and for constraining their redshifts using the 
photometric and, where available, the spectroscopic redshifts of the surrounding 
galaxies from the SDSS-DR8 data. A cluster was recognized when there were at 
least eight member galaxies within a radius of 560 kpc from the X-ray emission 
peak with photometric redshift in the redshift interval of the redshift of 
the likely identified BCG, $z_{\rm p,\,BCG} \pm 0.04(1+z_{\rm p,\,BCG})$.
The BCG was identified as the brightest galaxy among those galaxies within 
one arcmin from the X-ray position that show a peak in the histogram of 
their photometric redshifts.

The cluster photometric and spectroscopic redshift was measured as the 
weighted average of the photometric and the available spectroscopic 
redshifts, respectively, of the cluster galaxies within 560 kpc from 
the X-ray position. The measured redshifts agree well with the available 
redshifts in the literature; to date, 301 clusters are known as optically 
selected clusters with redshift measurements. Moreover, 310 clusters of 
the optically confirmed cluster sample have spectroscopic 
redshifts for at least one cluster member. The measured photometric redshifts 
agree well with the measured spectroscopic ones from the survey. 
The cluster redshifts of the optically confirmed cluster sample span  
a wide redshift range from 0.03 to 0.70. We reduced and analysed the X-ray 
data of this sample in an automated way to compute their X-ray properties. 

We presented a cluster catalogue from the survey comprising 345 X-ray-selected 
groups and clusters with their X-ray parameters derived from the 
spectral fits including the published sample in paper I. 
In addition to the best-fitting parameters, we estimated 
the physical properties ($R_{500}$, $L_{500}$ and $M_{500}$) of this sample
from an iterative procedure based on published scaling relations.
We investigated the \ltr relations for the first time based on a large 
cluster sample with X-ray spectroscopic parameters drawn from a single survey.
The current sample includes groups and clusters with wide ranges of 
temperatures and luminosities. The slope of the relation is consistent with 
the published ones of clusters with high temperatures and luminosities.
After excluding the low-luminosity  groups, we found no significant change 
of the slope and the intrinsic scatter of the relation with redshift when 
dividing the sample into three redshift bins. When including the low-luminosity  
groups in the low-redshift subsample, the slope was no longer consistent with the 
intermediate- and high-redshift subsamples.

In addition to the cluster sample with X-ray spectroscopic data, we presented the 
remainder of the optically confirmed cluster sample with their X-ray parameters 
based on the given flux in the 2XMMi-DR3 catalogue. We used 
the 2XMMi-DR3 flux because of their low-quality X-ray data, which are  
insufficient to perform spectral fitting. This sample comprises 185 groups 
and clusters with fluxes and luminosity in the energy band 0.5-2.0 keV 
and physical parameters ($R_{500}$, $L_{500}$, $M_{500}$, and $T_{500}$).

This is the largest X-ray-selected cluster catalogue to date based on  
XMM-Newton observations. It comprises 530 clusters with optical and 
X-ray properties, spanning the redshift range $0.03 < z < 0.70$. 
More than 75 percent of the cluster sample are newly discovered clusters 
in X-ray wavelengths. About 40 percent of the sample are systems new to the 
literature according to current entries in the NED.      

In the future we plan to study the remainder of the X-ray cluster candidates that  
were not detected by the current detection algorithm. They are either poor or 
at high redshifts. For the distant clusters, we plan follow-up by imaging and 
spectroscopy. For the X-ray cluster candidates with galaxies detected 
in SDSS imaging and have not been validated by the current algorithm, we 
plan to improve the current algorithm to constrain their redshifts. The new 
sample from the survey, especially the distant ones, will allow us to 
investigate the evolution of \ltr relation and X-ray-optical relations.    



 \begin{acknowledgements}
This project is supported by the Egyptian Ministry of Higher Education and 
Scientific Research (MHESR) in cooperation with the Leibniz-Institut f{\"u}r 
Astrophysik Potsdam (AIP), Germany. We acknowledge the partial support by 
the Deutsches Zentrum f{\"u}r Luft- und Raumfahrt (DLR) under contract number  
50 QR 0802. We also acknowledge financial support from the ARCHES project 
(7th Framework of the European Union, n$^{\circ}$ 313146). We thank the 
referee for the valuable comments that helped to 
improve the paper. The XMM-Newton project is an ESA Science Mission 
with instruments and contributions directly funded by ESA Member States and the 
USA (NASA). This research has made use of the NASA/IPAC Extragalactic Database 
(NED) which is operated by the Jet Propulsion Laboratory, California Institute 
of Technology, under contract with the National Aeronautics and Space 
Administration (NASA).
Funding for SDSS-III has been provided by the Alfred P. Sloan 
Foundation, the Participating Institutions, the National Science Foundation,
 and the U.S. Department of Energy. The SDSS-III web site is 
http://www.sdss3.org/. 
SDSS-III is managed by the Astrophysical Research 
Consortium for the Participating Institutions of the SDSS-III Collaboration 
including the University of Arizona, the Brazilian Participation Group, 
Brookhaven National Laboratory, University of Cambridge, University of Florida, 
the French Participation Group, the German Participation Group, the Instituto 
de Astrofisica de Canarias, the Michigan State/Notre Dame/JINA Participation 
Group, Johns Hopkins University, Lawrence Berkeley National Laboratory, Max 
Planck Institute for Astrophysics, New Mexico State University, New York 
University, Ohio State University, Pennsylvania State University, University of 
Portsmouth, Princeton University, the Spanish Participation Group, University 
of Tokyo, University of Utah, Vanderbilt University, University of Virginia, 
University of Washington, and Yale University.
 \end{acknowledgements}



\bibliographystyle{aa}
\bibliography{refbib2}

\addtocounter{table}{-3}
\clearpage
\begin{landscape}
\begin{table}
\caption{\label{tbl:SF_sample} First ten entries of the X-ray-selected 
group/cluster sample (345 objects) from the 2XMMi/SDSS Galaxy Cluster Survey 
with X-ray parameters from the spectral fitting.}
{\tiny
\begin{tabular}{c c c c c c c c c c c c c c c c c c}
\hline \hline
  \multicolumn{1}{c}{detid\tablefootmark{a}} &
  \multicolumn{1}{c}{Name\tablefootmark{a}} &
  \multicolumn{1}{c}{ra\tablefootmark{a}} &
  \multicolumn{1}{c}{dec\tablefootmark{a}} &
  \multicolumn{1}{c}{obsid\tablefootmark{a}} &
  \multicolumn{1}{c}{z\tablefootmark{b}} &
  \multicolumn{1}{c}{scale} &
  \multicolumn{1}{c}{$R_{\rm ap}$} &
  \multicolumn{1}{c}{$R_{\rm 500}$} &
  \multicolumn{1}{c}{$T_{\rm ap}$} &
  \multicolumn{1}{c}{$+eT_{\rm ap}$} &
  \multicolumn{1}{c}{$-eT_{\rm ap}$} &
  \multicolumn{1}{c}{$F_{\rm ap}$\tablefootmark{c}} &
  \multicolumn{1}{c}{$+eF_{\rm ap}$} &
  \multicolumn{1}{c}{$-eF_{\rm ap}$} &
  \multicolumn{1}{c}{$L_{\rm ap}$\tablefootmark{d}} &
  \multicolumn{1}{c}{$+eL_{\rm ap}$} &
  \multicolumn{1}{c}{$-eL_{\rm ap}$}  \\

  \multicolumn{1}{c}{} &
  \multicolumn{1}{c}{IAUNAME} &
  \multicolumn{1}{c}{(deg)} &
  \multicolumn{1}{c}{(deg)} &
  \multicolumn{1}{c}{} &
  \multicolumn{1}{c}{} &
  \multicolumn{1}{c}{kpc/$''$} &
  \multicolumn{1}{c}{(kpc)} &
  \multicolumn{1}{c}{(kpc)} &
  \multicolumn{1}{c}{(keV)} &
  \multicolumn{1}{c}{(keV)} &
  \multicolumn{1}{c}{(keV)} &
  \multicolumn{1}{c}{} &
  \multicolumn{1}{c}{} &
  \multicolumn{1}{c}{} &
  \multicolumn{1}{c}{} &
  \multicolumn{1}{c}{} &
  \multicolumn{1}{c}{} \\

(1)  &  (2)  &  (3)  & (4)  &   (5)   &  (6)  &   (7)   &   (8)   &  (9)   &  (10)    &  (11) &  (12) & (13) & (14) & (15) &  (16)  & (17)  &  (18)   \\
\hline 

002294 &     2XMM J001817.2+161740 &   4.57190 &  16.29470 & 0111000101 & 0.5401 & 6.35 & 476.50 &  810.94 & 4.57 & 0.78 & 0.60 &  16.74 & 0.58 & 0.79 & 144.72 &  5.58 &  5.52 \\
004444 &     2XMM J003318.4-212447 &   8.32687 & -21.41319 & 0044350101 & 0.1897 & 3.17 & 161.44 &  579.22 & 2.25 & 0.66 & 0.40 &   3.75 & 0.28 & 0.27 &   3.66 &  0.19 &  0.21 \\
005825 &     2XMM J003917.9+004200 &   9.82489 &   0.70013 & 0203690101 & 0.2801 & 4.24 & 152.81 &  483.64 & 1.43 & 0.77 & 0.29 &   0.88 & 0.06 & 0.06 &   2.10 &  0.13 &  0.17 \\
005842 &     2XMM J003922.4+004809 &   9.84343 &   0.80269 & 0203690101 & 0.4145 & 5.49 & 395.23 &  618.80 & 4.02 & 0.64 & 0.52 &   3.55 & 0.08 & 0.08 &  18.33 &  0.44 &  0.48 \\
005901 &     2XMM J003942.2+004533 &   9.92584 &   0.75919 & 0203690101 & 0.4156 & 5.50 & 247.44 &  589.18 & 2.35 & 0.43 & 0.33 &   2.51 & 0.12 & 0.08 &  14.02 &  0.67 &  0.55 \\
006070 &     2XMM J004039.2+253106 &  10.16344 &  25.51840 & 0153030101 & 0.1517 & 2.64 & 142.48 &  632.85 & 1.51 & 0.13 & 0.10 &  10.55 & 0.49 & 0.41 &   6.37 &  0.33 &  0.22 \\
006469 &     2XMM J004156.8+253151 &  10.48690 &  25.53105 & 0153030101 & 0.1278 & 2.28 & 150.73 &  579.30 & 3.18 & 1.09 & 0.77 &   5.21 & 0.19 & 0.34 &   2.10 &  0.10 &  0.09 \\
006920 &     2XMM J004231.2+005114 &  10.63008 &   0.85401 & 0090070201 & 0.1579 & 2.73 & 114.55 &  501.99 & 2.16 & 0.92 & 0.47 &   1.37 & 0.10 & 0.09 &   0.89 &  0.07 &  0.04 \\
007340 &     2XMM J004252.6+004259 &  10.71952 &   0.71650 & 0090070201 & 0.2697 & 4.13 & 421.41 &  579.12 & 3.12 & 0.90 & 0.61 &   4.14 & 0.19 & 0.15 &   8.45 &  0.44 &  0.34 \\
007362 &     2XMM J004253.7-093423 &  10.72397 &  -9.57311 & 0065140201 & 0.4069 & 5.43 & 260.60 &  613.30 & 3.29 & 1.25 & 0.74 &   3.03 & 0.20 & 0.24 &  15.14 &  1.16 &  0.94 \\

\hline
\end{tabular}
}
\end{table}


\addtocounter{table}{-1}
\begin{table}
\caption{\label{} continued.}
{\tiny
\begin{tabular}{c c c c c c c c c c c c c c c c}
\hline
\hline
  \multicolumn{1}{c}{detid\tablefootmark{a}} &
  \multicolumn{1}{c}{$L_{500}$\tablefootmark{e}} &
  \multicolumn{1}{c}{$\pm eL_{500}$} &
  \multicolumn{1}{c}{$M_{500}$\tablefootmark{f}} &
  \multicolumn{1}{c}{$\pm eM_{500}$} &
  \multicolumn{1}{c}{nH\tablefootmark{g}} &
  \multicolumn{1}{c}{objid\tablefootmark{h}} &
  \multicolumn{1}{c}{RA\tablefootmark{h}} &
  \multicolumn{1}{c}{Dec\tablefootmark{h}} &
  \multicolumn{1}{c}{$\bar{z}_{\rm p}$\tablefootmark{h}} &
  \multicolumn{1}{c}{$\bar{z}_{\rm s}$\tablefootmark{h}} &
  \multicolumn{1}{c}{$N_{z_{\rm s}}$\tablefootmark{h}} &
  \multicolumn{1}{c}{$z_{\rm type}$\tablefootmark{h}} &
  \multicolumn{1}{c}{offset\tablefootmark{h}} &
  \multicolumn{1}{c}{NED-Name} &
  \multicolumn{1}{c}{ref.}  \\

  \multicolumn{1}{c}{} &
  \multicolumn{1}{c}{} &
  \multicolumn{1}{c}{} &
  \multicolumn{1}{c}{} &
  \multicolumn{1}{c}{} &
  \multicolumn{1}{c}{} &
  \multicolumn{1}{c}{(BCG)} &
  \multicolumn{1}{c}{(deg)} &
  \multicolumn{1}{c}{(deg)} &
  \multicolumn{1}{c}{} &
  \multicolumn{1}{c}{} &
  \multicolumn{1}{c}{} &
  \multicolumn{1}{c}{} &
  \multicolumn{1}{c}{(kpc)} &
  \multicolumn{1}{c}{} &
  \multicolumn{1}{c}{}  \\

  (1)  &  (19)  & (20)  &  (21)  &  (22)   &  (23) &  (24)  &   (25)   &  (26) & (27) & (28)  & (29) & (30) & (31) & (32) & (33)   \\
\hline

002294 &  521.28 &  31.45 & 27.28 &  5.24 & 0.0393 & 1237679454926995783 &   4.57107 &  16.29433 & 0.5401 & 0.0000 &  0 & photo-z &  20.49 &                     RX J0018.2+1617 &                  1,2 \\
004444 &   17.49 &   0.71 &  6.67 &  1.35 & 0.0153 & 1237673016766496932 &   8.32630 & -21.41445 & 0.1897 & 0.0000 &  0 & photo-z &  17.22 &                                   - &                    - \\
005825 &    7.80 &   0.28 &  4.28 &  0.91 & 0.0198 & 1237663204918493337 &   9.82501 &   0.69981 & 0.2710 & 0.2801 &  1 &  spec-z &   5.14 &           SDSS CE J009.833157+00.701518 &                3,4,5 \\
005842 &   59.67 &   2.50 & 10.46 &  2.03 & 0.0197 & 1237663204918493446 &   9.84605 &   0.79222 & 0.3945 & 0.4145 &  3 &  spec-z & 213.01 &                                   - &                    - \\
005901 &   44.13 &   2.10 &  9.04 &  1.78 & 0.0195 & 1237663204918493223 &   9.92730 &   0.76163 & 0.3988 & 0.4156 &  2 &  spec-z &  56.24 &                WHL J003942.5+004541 &                  5,6 \\
006070 &   26.71 &   1.08 &  8.36 &  1.65 & 0.0368 & 1237678580906524886 &  10.16314 &  25.51779 & 0.1517 & 0.0000 &  0 & photo-z &   6.62 &                                   - &                    - \\
006469 &   14.21 &   0.07 &  6.26 &  1.27 & 0.0384 & 1237680071245365404 &  10.48821 &  25.52932 & 0.1278 & 0.0000 &  0 & photo-z &  18.10 &                                   - &                    - \\
006920 &    6.43 &   0.06 &  4.20 &  0.89 & 0.0179 & 1237663716555882709 &  10.63094 &   0.85020 & 0.1526 & 0.1579 &  4 &  spec-z &  36.87 &           GMBCG J010.63096+00.85021 &                  4,6 \\
007340 &   23.13 &   1.33 &  7.27 &  1.46 & 0.0178 & 1237663204918886547 &  10.71962 &   0.71844 & 0.2604 & 0.2697 &  4 &  spec-z &  28.79 &           SDSS CE J010.717058+00.725393 &                3,5,7 \\
007362 &   54.86 &   4.07 & 10.09 &  1.99 & 0.0270 & 1237652947993428384 &  10.72131 &  -9.57365 & 0.4069 & 0.0000 &  0 & photo-z &  52.83 &           GMBCG J010.72131-09.57365 &                  5,6 \\

\hline
\end{tabular}
}

\tablefoot{ The full catalogue is available at CDS.
\tablefoottext{a}{All these parameters are extracted from the 2XMMi-DR3 catalogue.} 
\tablefoottext{b}{The cluster redshift is from col. (28) otherwise from col. (27).}     
\tablefoottext{c}{Aperture X-ray flux $F_{\rm ap}$ [0.5-2.0] keV and its positive and negative errors in units of $10^{-14}$\ erg\ cm$^{-2}$\ s$^{-1}$.}
\tablefoottext{d}{Aperture X-ray luminosity $L_{\rm ap}$ [0.5-2.0] keV and its positive and negative errors in units of $10^{42}$\ erg\ s$^{-1}$.} 
\tablefoottext{e}{X-ray bolometric luminosity $L_{500}$ and its error in units of $10^{42}$\ erg\ s$^{-1}$.} 
\tablefoottext{f}{The cluster mass $M_{500}$ and its error  in units of $10^{13}$\ M$_\odot$.}
\tablefoottext{g}{The Galactic HI column in units $10^{22}$\ cm$^{-2}$.}
\tablefoottext{h}{These parameters are obtained from the developed optical detection algorithm.} 
   }

\tablebib{
1-  \cite{Romer00};       2-  \cite{Kolokotronis06};  3- \cite{Goto02}; 
4-  \cite{Koester07};     5-  \cite{Wen09};           6- \cite{Hao10}; 
7-  \cite{Plionis05};     8-  \cite{Merchan05};       9- \cite{Bahcall03}; 
10- \cite{Vikhlinin98};   11- \cite{Mullis03};       12- \cite{Gal03};  
13- \cite{Burenin07};     14- \cite{Horner08};       15- \cite{Finoguenov07};
16- \cite{McConnachie09}; 17- \cite{Olsen07};        18- \cite{Grove09}; 
19- \cite{Falco99};       20- \cite{Ramella01};      21- \cite{Zwicky61};  
22- \cite{dellAntonio94}; 23- \cite{Berlind06};      24- \cite{Dietrich07}; 
25- \cite{GUNN86};        26- \cite{Gladders05};     27- \cite{Barkhouse06};
28- \cite{McDowell03};    29- \cite{Schuecker04};    30- \cite{Wittman06}; 
31- \cite{Carlberg01};    32- \cite{Finoguenov09};   33- \cite{Hughes98};
34- \cite{Sehgal08};      35- \cite{Postman02}.     
}

\end{table}
\end{landscape}



\clearpage
\begin{landscape}
\begin{table}
\caption{\label{tbl:cat_sample} First ten entries of the X-ray-selected 
group/cluster sample (185 systems) from the 2XMMi/SDSS Galaxy Cluster Survey 
with X-ray parameters based on the given flux in the 2XMMi-DR3 catalogue.}
{\tiny
\begin{tabular}{c c c c c c c c c c c c c c c c c c}
\hline
\hline
  \multicolumn{1}{c}{detid\tablefootmark{a}} &
  \multicolumn{1}{c}{Name\tablefootmark{a}} &
  \multicolumn{1}{c}{ra\tablefootmark{a}} &
  \multicolumn{1}{c}{dec\tablefootmark{a}} &
  \multicolumn{1}{c}{obsid\tablefootmark{a}} &
  \multicolumn{1}{c}{z\tablefootmark{b}} &
  \multicolumn{1}{c}{scale} &
  \multicolumn{1}{c}{$R_{500}$} &
  \multicolumn{1}{c}{$F_{\rm cat}$\tablefootmark{a,c}} &
  \multicolumn{1}{c}{$\pm eF_{\rm cat}$} &
  \multicolumn{1}{c}{$L_{\rm cat}$\tablefootmark{d}} &
  \multicolumn{1}{c}{$\pm eL_{\rm cat}$} &
  \multicolumn{1}{c}{$L_{500}$\tablefootmark{e}} &
  \multicolumn{1}{c}{$\pm eL_{500}$} &
  \multicolumn{1}{c}{$M_{500}$\tablefootmark{f}} &
  \multicolumn{1}{c}{$\pm eM_{500}$} &
  \multicolumn{1}{c}{$T_{500}$} &
  \multicolumn{1}{c}{$\pm eT_{500}$} \\

  \multicolumn{1}{c}{} &
  \multicolumn{1}{c}{IAUNAME} &
  \multicolumn{1}{c}{(deg)} &
  \multicolumn{1}{c}{(deg)} &
  \multicolumn{1}{c}{} &
  \multicolumn{1}{c}{} &
  \multicolumn{1}{c}{kpc/$''$} &
  \multicolumn{1}{c}{(kpc)} &
  \multicolumn{1}{c}{} &
  \multicolumn{1}{c}{} &
  \multicolumn{1}{c}{} &
  \multicolumn{1}{c}{} &
  \multicolumn{1}{c}{} &
  \multicolumn{1}{c}{} &
  \multicolumn{1}{c}{} &
  \multicolumn{1}{c}{} &
  \multicolumn{1}{c}{(keV)}&
  \multicolumn{1}{c}{(keV)} \\

(1) &  (2)  &  (3)  & (4)  &   (5)   &  (6)  &   (7)  &   (8)  &  (9)  &  (10)  &  (11) &  (12) & (13) & (14) & (15) &  (16)  & (17)  &  (18) \\
\hline 

006511 &   2XMM J004205.5-093613 &  10.52296 &  -9.60375 & 0065140201 & 0.3256 & 4.71 &  582.39 &   3.30 & 0.47 &  11.51 &  1.63 &   29.35 &   4.66 &  7.87 &  1.67 & 1.84 & 0.45 \\
007481 &   2XMM J004259.7-092634 &  10.74900 &  -9.44286 & 0065140201 & 0.4151 & 5.49 &  678.59 &   7.97 & 1.21 &  49.11 &  7.44 &  106.34 &  20.98 & 13.80 &  2.93 & 2.65 & 0.63 \\
011071 &   2XMM J005608.0+004103 &  14.03365 &   0.68427 & 0303110401 & 0.4607 & 5.84 &  588.27 &   2.71 & 0.60 &  21.40 &  4.74 &   51.85 &  13.25 &  9.48 &  2.18 & 2.12 & 0.53 \\
014038 &   2XMM J010606.8+004926 &  16.52863 &   0.82407 & 0150870201 & 0.2564 & 3.98 &  680.44 &  13.69 & 1.54 &  27.55 &  3.11 &   60.33 &   7.82 & 11.62 &  2.34 & 2.30 & 0.55 \\
014050 &   2XMM J010610.0+005110 &  16.54201 &   0.85302 & 0150870201 & 0.2566 & 3.99 &  689.89 &  15.44 & 1.42 &  31.12 &  2.86 &   65.80 &   7.60 & 12.12 &  2.41 & 2.36 & 0.56 \\
021043 &   2XMM J015558.5+053159 &  28.99394 &   5.53329 & 0153030701 & 0.4312 & 5.62 &  671.20 &   5.82 & 0.85 &  39.27 &  5.76 &  105.58 &  20.88 & 13.61 &  2.90 & 2.64 & 0.63 \\
021688 &   2XMM J020056.5-092119 &  30.23615 &  -9.35526 & 0203840201 & 0.3381 & 4.83 &  549.44 &   2.29 & 0.24 &   8.72 &  0.92 &   21.37 &   2.51 &  6.70 &  1.40 & 1.67 & 0.41 \\
023255 &   2XMM J021447.5-005425 &  33.69817 &  -0.90720 & 0201020201 & 0.2650 & 4.08 &  484.82 &   0.88 & 0.12 &   1.91 &  0.27 &    7.50 &   0.78 &  4.24 &  0.93 & 1.23 & 0.32 \\
030889 &   2XMM J023458.7-085055 &  38.74463 &  -8.84868 & 0150470601 & 0.2484 & 3.89 &  603.14 &   4.15 & 0.53 &   7.77 &  1.00 &   27.63 &   4.52 &  8.02 &  1.70 & 1.83 & 0.45 \\
033092 &   2XMM J024810.2+311511 &  42.04268 &  31.25311 & 0111490401 & 0.3871 & 5.27 &  532.12 &   2.17 & 0.39 &  11.31 &  2.03 &   20.99 &   4.49 &  6.44 &  1.46 & 1.64 & 0.42 \\
034341 &   2XMM J030212.0+001108 &  45.55036 &   0.18583 & 0041170101 & 0.6523 & 6.94 &  413.91 &   0.43 & 0.08 &   7.91 &  1.49 &   12.12 &   2.98 &  4.15 &  1.01 & 1.33 & 0.35 \\

\hline
\end{tabular}
}

\end{table}


\addtocounter{table}{-1}
\begin{table}
\caption{\label{} continued.}
{\tiny
\begin{tabular}{c c c c c c c c c c c}
\hline
\hline
  \multicolumn{1}{c}{detid\tablefootmark{a}} &
  \multicolumn{1}{c}{objid\tablefootmark{g}} &
  \multicolumn{1}{c}{RA\tablefootmark{g}} &
  \multicolumn{1}{c}{DEC\tablefootmark{g}} &
  \multicolumn{1}{c}{$\bar{z}_{\rm p}$\tablefootmark{g}} &
  \multicolumn{1}{c}{$\bar{z}_{\rm s}$\tablefootmark{g}} &
  \multicolumn{1}{c}{$N_{z_{\rm s}}$\tablefootmark{g}} &
  \multicolumn{1}{c}{$z_{\rm type}$\tablefootmark{g}} &
  \multicolumn{1}{c}{offset\tablefootmark{g}} &
  \multicolumn{1}{c}{NED-Name} &
  \multicolumn{1}{c}{ref.}  \\

  \multicolumn{1}{c}{}&
  \multicolumn{1}{c}{(BCG)} &
  \multicolumn{1}{c}{(deg)} &
  \multicolumn{1}{c}{(deg)} &
  \multicolumn{1}{c}{} &
  \multicolumn{1}{c}{} &
  \multicolumn{1}{c}{} &
  \multicolumn{1}{c}{} &
  \multicolumn{1}{c}{(kpc)} &
  \multicolumn{1}{c}{} &
  \multicolumn{1}{c}{}  \\

  (1)  &  (19)  & (20)  &  (21)  &  (22)   &  (23) &  (24)  &   (25)   &  (26) & (27) & (28)  \\
\hline

006511 & 1237652947993297563 &  10.51514 &  -9.60060 & 0.3256 & 0.0000 &  0 & photo-z & 138.56 &                                   - &                    - \\
007481 & 1237652630713795354 &  10.75138 &  -9.43350 & 0.4151 & 0.0000 &  0 & photo-z & 191.29 &                CXO J004259.9-092704 &                    1 \\
011071 & 1237663204920328298 &  14.04250 &   0.68188 & 0.4607 & 0.0000 &  0 & photo-z & 192.67 &                                   - &                    - \\
014038 & 1237663204921376994 &  16.52926 &   0.81949 & 0.2515 & 0.2564 &  4 &  spec-z &  68.21 &       SDSS CE J016.528793+00.817471 &            2,3,4,5,6 \\
014050 & 1237663785278374092 &  16.54324 &   0.85569 & 0.2492 & 0.2566 &  5 &  spec-z &  42.90 &          MaxBCG J016.54324+00.85569 &                4,7,8 \\
021043 & 1237678663047250389 &  28.98754 &   5.53073 & 0.4312 & 0.0000 &  0 & photo-z & 143.57 &                                   - &                    - \\
021688 & 1237652900224303421 &  30.23274 &  -9.35660 & 0.3472 & 0.3381 &  1 &  spec-z &  62.93 &                                   - &                    - \\
023255 & 1237680000377684204 &  33.69949 &  -0.90894 & 0.2649 & 0.2650 &  1 &  spec-z &  32.01 &                                   - &                    - \\
030889 & 1237653500970270877 &  38.74547 &  -8.84926 & 0.2484 & 0.0000 &  0 & photo-z &  13.80 &                                   - &                    - \\
033092 & 1237670458043073373 &  42.04490 &  31.25411 & 0.3871 & 0.0000 &  0 & photo-z &  39.90 &                                   - &                    - \\
034341 & 1237663784217346252 &  45.54822 &   0.18751 & 0.6516 & 0.6523 &  1 &  spec-z &  67.99 &                 BLOX J0302.2+0010.5 &                    9 \\

\hline
\end{tabular}
}

\tablefoot{ The full catalogue is available at CDS and contains the information given in columns (1)-(28) in Table~\ref{tbl:cat_sample}.
\tablefoottext{a}{All these parameters are extracted from the 2XMMi-DR3 catalogue.} 
\tablefoottext{b}{The cluster redshift is from col. (23) otherwise from col. (22).}     
\tablefoottext{c}{The given flux in the 2XMMi-DR3 $F_{\rm cat}$ [0.5-2.0] keV and its errors in units of $10^{-14}$\ erg\ cm$^{-2}$\ s$^{-1}$.}
\tablefoottext{d}{The computed X-ray luminosity $L_{\rm cat}$ [0.5-2.0] keV and its errors in units of $10^{42}$\ erg\ s$^{-1}$.} 
\tablefoottext{e}{X-ray bolometric luminosity $L_{500}$ and its error in units of $10^{42}$\ erg\ s$^{-1}$.} 
\tablefoottext{f}{The cluster mass $M_{500}$ and its error  in units of $10^{13}$\ M$_\odot$.}
\tablefoottext{g}{These parameters are obtained from our detection algorithm in the optical band.}
}

\tablebib{
1-  \cite{Evans10};        2-  \cite{Goto02};      3-  \cite{Lopes04};
4-  \cite{Barkhouse06};    5-  \cite{Wen09};       6-  \cite{Hao10};
7-  \cite{Koester07};      8-  \cite{Bahcall03};   9-  \cite{Dietrich07};
10- \cite{GUNN86};         11- \cite{Knobel09};    12- \cite{Olsen07};
13- \cite{Finoguenov07};   14- \cite{Gal03};       15- \cite{McConnachie09};
16- \cite{Merchan05};      17- \cite{Boschin02};   18- \cite{Kolokotronis06}; 
19- \cite{Horner08};       20- \cite{Zwicky61};    21- \cite{Falco99};
22- \cite{Burenin07};      23-  \cite{Yoon08};     24- \cite{Romer00};
25- \cite{White99}.
}

\end{table}
\end{landscape}


\end{document}